\documentclass[sigconf]{acmart}
\usepackage{pgfplots}\pgfplotsset{compat=1.18}
\usepackage{pgfplotstable}
\usepackage{multirow}

\newcommand{\sampleN}{81}
\newcommand{\nTrials}{648}

\newcommand{\changeRefute}{45.7\%}
\newcommand{\changeConfirm}{1.2\%}

\newcommand{\changeRefutePart}{79\%}
\newcommand{\changeConfirmPart}{4.9\%}
\newcommand{\mcnemarChi}{58.0}
\newcommand{\mcnemarP}{<.001}

\newcommand{\confPriorRefute}{4.42}
\newcommand{\confAfterRefute}{4.34}
\newcommand{\confPriorConfirm}{4.46}
\newcommand{\confAfterConfirm}{5.42}

\newcommand{\rtConfirm}{2.7}
\newcommand{\rtRefute}{6.0}
\newcommand{\rtConfirmModel}{2.2}
\newcommand{\rtRefuteModel}{3.6}
\newcommand{\rtFasterPct}{39\%}

\newcommand{\pickAiP}{<.001}

\newcommand{\pickVoiceP}{=.41}

\newcommand{\pickIntP}{=.28}

\newcommand{\shiftDiff}{1.05}
\newcommand{\shiftConfirm}{0.96}
\newcommand{\shiftRefute}{-0.09}
\newcommand{\shiftAiP}{<.001}
\newcommand{\shiftVoiceP}{=.85}
\newcommand{\shiftIntP}{=.91}

\newcommand{\rtAiP}{<.001}
\newcommand{\rtVoiceP}{=.64}
\newcommand{\rtIntP}{=.94}

\newcommand{\friedmanChi}{88.2}
\newcommand{\friedmanP}{<.001}
\newcommand{\kendallW}{0.36}

\newcommand{\demoMalePct}{51.9}

\newcommand{\demoFemPct}{44.4}
\newcommand{\demoNbN}{3}
\newcommand{\demoNbPct}{3.7}

\newcommand{\demoAgeYoungPct}{53.1}

\newcommand{\demoAgeMidPct}{32.1}

\newcommand{\demoAgeOldPct}{14.8}
\newcommand{\demoUnderThirtyfive}{85}
\newcommand{\demoAnthroFirstN}{39}
\newcommand{\demoAnthroFirstPct}{48.1}
\newcommand{\demoMachineFirstN}{42}
\newcommand{\demoMachineFirstPct}{51.9}
\newcommand{\expScMed}{3}

\newcommand{\expMean}{3.5}

\newcommand{\shapiroShiftW}{0.975}
\newcommand{\shapiroShiftP}{<.001}
\newcommand{\shapiroRtW}{0.896}
\newcommand{\shapiroRtP}{<.001}

\newcommand{\stayerPct}{76.5}
\newcommand{\stayerShiftDiff}{1.12}
\newcommand{\stayerShiftP}{<.001}

\newcommand{\confirmChangeN}{4}
\newcommand{\confirmChangeTotalN}{324}

\providecommand{\pickOR}{}\renewcommand{\pickOR}{230}
\providecommand{\pickORlo}{}\renewcommand{\pickORlo}{61}
\providecommand{\pickORhi}{}\renewcommand{\pickORhi}{865}
\providecommand{\pickAiP}{}\renewcommand{\pickAiP}{<.001}
\providecommand{\pickIntRatio}{}\renewcommand{\pickIntRatio}{3.66}
\providecommand{\pickIntRatioLo}{}\renewcommand{\pickIntRatioLo}{0.34}
\providecommand{\pickIntRatioHi}{}\renewcommand{\pickIntRatioHi}{38.99}
\providecommand{\pickIntP}{}\renewcommand{\pickIntP}{=.282}
\providecommand{\pickVoiceP}{}\renewcommand{\pickVoiceP}{=.410}

\providecommand{\shiftEff}{}\renewcommand{\shiftEff}{1.01}
\providecommand{\shiftEffLo}{}\renewcommand{\shiftEffLo}{0.87}
\providecommand{\shiftEffHi}{}\renewcommand{\shiftEffHi}{1.15}
\providecommand{\shiftAiP}{}\renewcommand{\shiftAiP}{<.001}
\providecommand{\shiftInt}{}\renewcommand{\shiftInt}{-0.05}
\providecommand{\shiftIntLo}{}\renewcommand{\shiftIntLo}{-0.33}
\providecommand{\shiftIntHi}{}\renewcommand{\shiftIntHi}{0.24}
\providecommand{\shiftIntP}{}\renewcommand{\shiftIntP}{=.759}
\providecommand{\shiftVoiceP}{}\renewcommand{\shiftVoiceP}{=.928}

\providecommand{\rtRatio}{}\renewcommand{\rtRatio}{1.82}
\providecommand{\rtRatioLo}{}\renewcommand{\rtRatioLo}{1.64}
\providecommand{\rtRatioHi}{}\renewcommand{\rtRatioHi}{2.02}
\providecommand{\rtAiP}{}\renewcommand{\rtAiP}{<.001}
\providecommand{\rtIntRatio}{}\renewcommand{\rtIntRatio}{0.92}
\providecommand{\rtIntRatioLo}{}\renewcommand{\rtIntRatioLo}{0.75}
\providecommand{\rtIntRatioHi}{}\renewcommand{\rtIntRatioHi}{1.13}
\providecommand{\rtIntP}{}\renewcommand{\rtIntP}{=.420}
\providecommand{\rtVoiceP}{}\renewcommand{\rtVoiceP}{=.533}
\providecommand{\rtConfirmModel}{}\renewcommand{\rtConfirmModel}{2.6}
\providecommand{\rtRefuteModel}{}\renewcommand{\rtRefuteModel}{4.7}
\providecommand{\rtFasterPct}{}\renewcommand{\rtFasterPct}{45\%}
\providecommand{\rtLongN}{}\renewcommand{\rtLongN}{7}
\providecommand{\rtMaxS}{}\renewcommand{\rtMaxS}{172}
\providecommand{\smallestVoiceP}{}\renewcommand{\smallestVoiceP}{=.282}
\providecommand{\stayerPct}{}\renewcommand{\stayerPct}{76.5}
\providecommand{\stayerShiftDiff}{}\renewcommand{\stayerShiftDiff}{1.08}
\providecommand{\stayerShiftP}{}\renewcommand{\stayerShiftP}{<.001}
\providecommand{\refStayN}{}\renewcommand{\refStayN}{176}
\providecommand{\refSwitchN}{}\renewcommand{\refSwitchN}{148}
\providecommand{\refStayShift}{}\renewcommand{\refStayShift}{-0.36}
\providecommand{\refSwitchShift}{}\renewcommand{\refSwitchShift}{+0.24}
\providecommand{\friedmanPickChi}{}\renewcommand{\friedmanPickChi}{123.9}
\providecommand{\friedmanPickP}{}\renewcommand{\friedmanPickP}{<.001}
\providecommand{\kendallPickW}{}\renewcommand{\kendallPickW}{0.51}
\providecommand{\friedmanChi}{}\renewcommand{\friedmanChi}{88.2}
\providecommand{\friedmanP}{}\renewcommand{\friedmanP}{<.001}
\providecommand{\kendallW}{}\renewcommand{\kendallW}{0.36}
\providecommand{\posthocAiMaxP}{}\renewcommand{\posthocAiMaxP}{<.001}
\providecommand{\posthocVoiceMinP}{}\renewcommand{\posthocVoiceMinP}{=.847}
\providecommand{\mcnemarChi}{}\renewcommand{\mcnemarChi}{58.0}
\providecommand{\mcnemarP}{}\renewcommand{\mcnemarP}{<.001}
\providecommand{\changeRefutePart}{}\renewcommand{\changeRefutePart}{79\%}
\providecommand{\changeConfirmPart}{}\renewcommand{\changeConfirmPart}{4.9\%}
\providecommand{\shapiroShiftW}{}\renewcommand{\shapiroShiftW}{0.975}
\providecommand{\shapiroShiftP}{}\renewcommand{\shapiroShiftP}{<.001}
\providecommand{\shapiroRtW}{}\renewcommand{\shapiroRtW}{0.896}
\providecommand{\shapiroRtP}{}\renewcommand{\shapiroRtP}{<.001}

\usetikzlibrary{positioning, arrows.meta, calc, shapes.geometric}
\usepgfplotslibrary{groupplots}

\definecolor{cConfirm}{HTML}{2A9D8F} 
\definecolor{cRefute}{HTML}{E76F51}  
\definecolor{cAnthro}{HTML}{4C72B0}  
\definecolor{cMachine}{HTML}{9A9A9A} 
\definecolor{cBar}{HTML}{4C72B0}     

\pgfplotsset{
  studbase/.style={
    width=\columnwidth, height=4.7cm,
    ybar, bar width=11pt,
    ymajorgrids, grid style={gray!20, thin},
    axis line style={gray!55}, tick style={gray!55},
    tick align=outside,
    every axis label/.append style={font=\small},
    tick label style={font=\small},
    enlarge x limits=0.45,
    legend cell align=left,
    legend style={draw=none, fill=none, font=\small,
      at={(0.5,1.02)}, anchor=south, legend columns=-1,
      /tikz/every even column/.append style={column sep=10pt}},
  },
  studerr/.style={
    error bars/.cd, y dir=both, y explicit,
    error bar style={black!65, line width=0.6pt}, error mark options={black!65},
  },
  studval/.style={
    forget plot, draw=none, fill=none, point meta=explicit,
    nodes near coords,
    every node near coord/.append style={
      anchor=south, font=\scriptsize, text=black!80, inner sep=1.6pt, yshift=1.5pt},
  },
  studsig/.style={},
}

\acmConference[STUD SS2026]{Student Research Project}{Summer Semester~2026}{TH Ingolstadt, Germany}
\acmYear{2026}
\copyrightyear{2026}
\acmDOI{}
\acmISBN{}
\setcopyright{none}

\begin{document}

\title[What's the Difference? Voice Design and Confirmation Bias in AI Self-Diagnosis]{What's the Difference?}
\subtitle{Voice Design and Confirmation Bias in AI-Assisted Self-Diagnosis: Anthropomorphic vs.\ Machine-Like Voice Output}

\author{Niklas R\"othlingsh\"ofer}
\authornote{The first three authors contributed equally to this research.}
\affiliation{\institution{Technische Hochschule Ingolstadt}\city{Ingolstadt}\country{Germany}}
\author{Emma Winkler}
\authornotemark[1]
\affiliation{\institution{Technische Hochschule Ingolstadt}\city{Ingolstadt}\country{Germany}}
\author{Melanie L\"offler}
\authornotemark[1]
\affiliation{\institution{Technische Hochschule Ingolstadt}\city{Ingolstadt}\country{Germany}}
\author{Emely Rosbach}
\affiliation{\institution{Technische Hochschule Ingolstadt}\city{Ingolstadt}\country{Germany}}

\renewcommand{\shortauthors}{R\"othlingsh\"ofer et al.}

\begin{abstract}
As symptom checkers based on artificial intelligence (AI) are becoming common entry points
for health information, their influence on lay users' diagnostic decisions raises questions
about trust and bias. Of particular concern is confirmation bias, the tendency to favour
information that supports one's existing belief while discounting information that
contradicts it. One of the central design choices for such systems is the style of voice:
whether they should present themselves with a human-like (anthropomorphic) or a clearly
machine-like (xenomorphic) voice. This study investigates
whether the voice design of an AI symptom checker modulates confirmation bias in
self-diagnosis. In a within-subjects online experiment ($N=\sampleN$ participants,
\nTrials{} trials),
participants selected a diagnosis and rated their confidence, then heard an AI-generated
recommendation that either confirmed or contradicted their initial choice, delivered by
an anthropomorphic or a machine-like voice, before deciding again. We observed a strong,
direction-dependent influence of the AI: participants revised their diagnosis in
\changeRefute{} of trials when the AI contradicted them versus \changeConfirm{} when it
confirmed them (generalised linear mixed model, GLMM, $p\pickAiP$); confidence rose under
confirmation but barely moved under contradiction (robust linear mixed model, $p\shiftAiP$); and decisions took longer after contradiction
(Gamma GLMM, $p\rtAiP$). This effect did not differ between the two voices (no voice or
interaction term was significant, smallest $p\smallestVoiceP$), even though a manipulation check confirmed the voices were
perceived as markedly different. The findings suggest that the \emph{direction} of an AI
recommendation, rather than the anthropomorphism of its voice, drives confirmation-related
bias in AI-assisted self-diagnosis.

\end{abstract}



\maketitle
\hypersetup{pdftitle={What's the Difference? Voice Design and Confirmation Bias in AI-Assisted Self-Diagnosis: Anthropomorphic vs. Machine-Like Voice Output}}

\section{Introduction}

Symptom checkers based on artificial intelligence (AI) are increasingly used as accessible
tools for preliminary health assessment. By allowing users to enter symptoms and receive
possible diagnoses or triage advice, these systems may support early orientation before
professional consultation. However, prior research has raised concerns about their
diagnostic reliability, safety, and suitability for laymen~\citep{wallace2022,mentzou2025}.

In medical decision contexts, the interaction between human judgement and AI
recommendation is particularly sensitive. Users may not only benefit from AI support but
also become vulnerable to automation bias, relying too strongly on automated
recommendations even when they are wrong~\citep{kuecking2024,qazi2026,rosbach2025automation}.

Such reliance is further shaped by cognitive biases, that is, systematic deviations from
rational judgement that arise from the way people process information. One of the most
studied biases is \emph{confirmation bias}, the tendency to seek out, weigh more heavily
(i.e., perceive as higher in quality), and more readily accept information that is consistent with one's prior beliefs while discounting information that contradicts
them~\citep{nickerson1998}. In self-diagnosis contexts this bias is particularly relevant:
users typically form an initial assumption about their symptoms, and they may therefore be
more likely to accept AI feedback that supports that assumption than feedback that
challenges it. This carries a concrete risk: when both the user and the AI are wrong, an
agreeing recommendation can produce a \emph{false confirmation}~\citep{rosbach2024}. A
layperson may, for instance, be falsely reassured that their symptoms do not require
immediate medical attention and therefore fail to seek help in time.

Voice-based AI assistants add another layer to this interaction: spoken recommendations are
not perceived as neutral information, since vocal characteristics shape perceived
credibility, social presence, and trust~\citep{seaborn2021}, and trust in healthcare voice
assistants has been shown to depend on perceived usefulness, content credibility, service
quality, and risk-related factors~\citep{zhan2024}. This raises a central design question: whether patient-facing AI
should imitate human interaction or present itself as clearly non-human. Anthropomorphic
design is often pursued precisely to make a system feel familiar and to engender trust; a
machine-like (xenomorphic) design instead emphasises the system's artificial nature and may
reduce patient confusion by clearly distinguishing AI from human
physicians~\citep{milford2025}.

While previous work has examined the accuracy of symptom checkers and trust in healthcare
voice assistants, less is known about how the voice design of an AI symptom checker
influences users' diagnostic decisions after confirming or contradicting advice. This study
addresses that gap, testing whether users exhibit a general confirmation bias toward the AI -
across diagnostic choice, confidence, and response time - and, as its primary question,
whether the strength of that bias is moderated by the AI's voice design (anthropomorphic
vs.\ machine-like). The formal hypotheses are stated in the Methods section.

This paper makes three contributions: (i) a controlled within-subjects design that
isolates the direction of an AI recommendation from its vocal presentation in a
self-diagnosis task; (ii) mixed-effects evidence for a strong, direction-dependent
confirmation bias across choice, confidence, and response time; and (iii) the finding
that this bias is unaffected by whether the AI speaks with an anthropomorphic or a
machine-like voice, despite a successful perceptual manipulation.

\section{Related Work}

\subsection{Confirmation Bias in AI-Assisted Decision-Making}
In his comprehensive
review, \citet{nickerson1998} characterises confirmation bias as a pervasive and largely unintentional
feature of everyday reasoning rather than a deliberate strategy, spanning biased search
for information, biased interpretation of ambiguous evidence, and biased memory.

The strength with which the bias operates is not fixed but modulated by several factors.
Four conditions in particular make it stronger: a firm prior commitment to the initial
belief, which raises the psychological stakes of maintaining it; evidence that is
ambiguous enough to be read as supportive; a source perceived as credible or trustworthy;
and a personal motivation to reach a particular conclusion~\citep{nickerson1998}. These moderators matter directly for
AI advice: an AI recommendation is an external piece of evidence that a user must
integrate with an already-formed belief, and its perceived credibility and the ambiguity
of the underlying symptoms determine how readily it is accepted or resisted.

In AI-assisted self-diagnosis this bias is particularly relevant, because users often
enter the interaction with an initial assumption about their symptoms. When an AI
recommendation agrees with that assumption, users may become more confident in their
initial diagnosis even if alternative explanations remain plausible; a contradicting
recommendation may instead be discounted because it conflicts with a prior belief.
Confirmation bias thus provides a theoretical basis for studying how users react to
confirming versus contradicting AI advice.

Empirically, this bias has been observed in each of the three outcomes we examine. For
\emph{choices}, mental-health professionals and students who had first made a preliminary
diagnosis of a clinical vignette rated a mock AI triage recommendation as more trustworthy and
were more likely to accept it when it matched that diagnosis~\citep{bashkirova2024}. For
\emph{confidence}, agreeing advice raises confidence even when it barely changes the judgement
itself: in judge--advisor studies with numeric estimates, advice close to one's initial
estimate acts as social validation that increases confidence, an effect that shrinks and
eventually vanishes as advice moves further away~\citep{schultze2015}. People place
confidence even in a spurious consensus of redundant, agreeing opinions, whereas they
discount independent opinions that conflict with one another, so that confidence and
accuracy dissociate~\citep{yaniv2009}; and when participants could adjust an incentivised confidence
wager after seeing a partner's judgement, agreement raised confidence more than disagreement
lowered it, and participants took the strength of confirming but not of disconfirming
opinions into account~\citep{kappes2020}. Comparable agreement effects occur with AI advice:
advice that agreed with participants' initial judgement led to larger increases in their
certainty, and once heeded it was used similarly whether it was labelled as coming from an AI
or from peers~\citep{vodrahalli2022}. \emph{Decision time} has been studied less directly.
Conflicting information can prolong decisions - decision makers spent more time per decision
when a peer's second opinion frequently disagreed with the AI's recommendation~\citep{lu2024} -
and time constraints increase anchoring on AI advice: participants given 25 rather than 10
seconds disagreed more often with incorrect AI predictions~\citep{rastogi2022}, consistent with
fast decisions reflecting less deliberate processing of the advice. Together, these findings
lead us to expect that contradicting AI advice triggers more revisions, a weaker increase in
confidence, and longer deliberation than confirming advice.

\subsection{AI-Based Self-Diagnosis and Healthcare Chatbots}
AI-based self-diagnosis tools and healthcare chatbots are increasingly used as accessible,
immediate sources of health information and preliminary guidance. \citet{yu2024}, for
example,
examine the acceptance of healthcare chatbots among older adults with an extended UTAUT
model and show that performance expectancy, effort expectancy, and social influence shape
the intention to use such systems. At
the same time, using AI for self-diagnosis introduces risks. \citet{zada2025}
investigate medical misinformation in AI-assisted self-diagnosis and find that
AI-generated medical advice is often unclear or inaccurate, so that relying solely on such
systems for self-diagnosis increases the risk of being misinformed.

This lack of expertise is what makes medical decision-making especially difficult for
laypeople. Unlike clinicians, lay users have no formal training to interpret symptoms,
to weigh competing differential diagnoses against one another, or to judge how reliable a
given AI output is for their particular case~\citep{mentzou2025,you2021}. As a result they
have few independent grounds on which to check a recommendation, which makes them more
dependent on - and correspondingly more swayable by - whatever the system suggests, and
less able to detect when its advice is wrong - a concern given that such advice is often
unclear or inaccurate~\citep{zada2025}.

\subsection{Accuracy and Safety of AI Symptom Checkers}
The safety of AI-based self-diagnosis depends on more than technical performance; it also
depends on how users interpret and act on system outputs. \citet{mentzou2025}
review digital self-diagnosis tools and highlight important gaps in their development,
impact, and oversight, so accuracy and safety cannot be judged through diagnostic
correctness alone. Capability is rising: \citet{tu2025} present AMIE, a
conversational diagnostic AI with strong performance in simulated consultations, yet such
controlled results do not represent everyday symptom-checker use by laymen.
\citet{you2021} further found that chatbot-based symptom checkers often lack support for
comprehensive histories, flexible input, and diverse users, and warn that users may be at
risk if they blindly trust the results. The design of the interaction itself may thus
influence whether users understand, trust, and follow AI-generated advice.

\subsection{Biases in AI-Assisted Medical Decision-Making}
Beyond accuracy, AI recommendations can shape human decisions. Automation bias refers to
the tendency to rely too strongly on automated recommendations even when they may be
incorrect. \citet{kuecking2024} studied automation bias in AI-based
clinical decision support and found that non-specialists were especially susceptible to
accepting incorrect AI recommendations. We expect this susceptibility to be even more
pronounced in the everyday use of symptom checkers by laypeople. Preliminary evidence from
\citet{qazi2026} suggests that even medically trained, AI-literate users can be affected by
erroneous large-language-model recommendations in diagnostic reasoning.

Closely related to the present study, \citet{rosbach2024} show - in computational
pathology - a direct confirmation-bias effect of AI advice: when an AI recommendation
mirrored a pathologist's own flawed tumour-percentage estimate, that erroneous estimate was
reinforced rather than corrected, an effect the authors term \emph{false confirmation}, and
one that time pressure, counter-intuitively, weakened rather than worsened. This maps
directly onto the situation we study, in a lay self-diagnosis rather than an expert pathology
setting: an AI recommendation that agrees with a user's initial diagnosis, whether right or
wrong, may entrench it in the same way, which is exactly the mechanism our confirming
condition isolates - underscoring, together with broader analyses of bias in medical
AI~\citep{cross2024}, that the risks of AI-assisted diagnosis depend not just on system
accuracy but on how users respond when an answer confirms or contradicts their own judgement.

\subsection{Voice Design, Anthropomorphism, and Machine-Like AI}
How AI advice is presented may further affect perception and decision-making. In
voice-based systems the recommendation is processed as speech and not merely as text, and
vocal characteristics can influence how confident, credible, or socially present a system
appears. \citet{kirkland2022} show that speech features
such as filled pauses, speech rate, and fundamental frequency affect perceived speaker
confidence, suggesting that users read acoustic properties as cues to certainty or
reliability. More broadly, a survey of voice in human--agent interaction reports that
human-like voices are generally rated as more likeable and often as more trustworthy or
credible than synthetic ones, whereas evidence for behavioural consequences such as
persuasiveness is inconsistent~\citep{seaborn2021}. Voice design may therefore affect decisions independently of content: \citet{dubiel2024} found that synthetic voice quality influenced participants'
choices even when the available options were held constant, and frame this as a potential
dark pattern. A related question is whether medical AI should appear human-like or clearly
non-human. \citet{milford2025} argue against anthropomorphism in
patient-facing medical AI and advocate a \emph{xenomorphic} approach that emphasises
non-human characteristics to distinguish AI from human physicians, arguing that
human-like designs built on trust-engendering features fall into the uncanny valley,
import cultural biases about who appears trustworthy, and risk competing with the human
doctor--patient relationship. Beyond being a mere presentation choice, anthropomorphism
may thus itself act as a bias: attributing human traits, credibility, or agency to a
system can shape trust independently of the actual content of its advice. Such attributions are not neutral, either:
anthropomorphic conversational assistants are frequently gendered: even where companies
claim otherwise, system outputs are ambiguous about their humanness and users personify and
gender them~\citep{abercrombie2021}. This
distinction matters in healthcare, where human-like systems may increase trust but also
blur the boundary between automated advice and professional medical judgement. In line with
this literature we operationalise the non-human design as a \emph{machine-like} voice and
contrast it with an anthropomorphic one.

These strands - confirmation and automation bias in AI-assisted diagnosis, the particular
vulnerability of laypeople, and the perceptual effects of voice design - have rarely been
combined, leaving open whether the voice of an AI symptom checker modulates how strongly
users maintain or revise an initial diagnosis after confirming or contradicting advice: the
question the present study addresses.

\section{Methods}

\newcommand{\stepttl}[1]{{\small\bfseries #1}}
\newcommand{\stepsub}[1]{{\scriptsize\textcolor{black!55}{#1}}}
\begin{figure*}
  \centering
  \begin{tikzpicture}[
      font=\small,
      box/.style={rounded corners=2.5pt, draw=cAnthro!45, fill=cAnthro!7,
                  text width=3.0cm, align=center, minimum height=0.8cm, inner sep=3pt},
      hl/.style ={fill=cAnthro!20, draw=cAnthro!70},
      ai/.style ={fill=cConfirm!12, draw=cConfirm!60!black},
      arr/.style={-{Stealth[length=5pt]}, draw=cAnthro!60, line width=0.9pt},
    ]
    \node[box]                       (d)  {\stepttl{1~~Demographics}\\[1pt]\stepsub{age, gender, experience}};
    \node[box, right=0.7cm of d]     (pr) {\stepttl{2~~Practice}\\[1pt]\stepsub{warm-up trial}};
    \node[box, hl, right=0.7cm of pr](st) {\stepttl{3~~Main study}\\[1pt]\stepsub{8 trials, 2 voice blocks}};
    \node[box, right=0.7cm of st]    (cl) {\stepttl{4~~Closing}\\[1pt]\stepsub{voice ratings}};
    \draw[arr] (d) -- (pr);  \draw[arr] (pr) -- (st);  \draw[arr] (st) -- (cl);
    \node[box, below=1.05cm of d]     (v)   {\stepttl{Symptom vignette}\\[1pt]\stepsub{3 diagnosis options}};
    \node[box, right=0.7cm of v]      (ip)  {\stepttl{Initial decision}\\[1pt]\stepsub{+ confidence (1--7)}};
    \node[box, ai, right=0.7cm of ip] (rec) {\stepttl{AI recommendation}\\[1pt]\stepsub{confirm / contradict}};
    \node[box, right=0.7cm of rec]    (fp)  {\stepttl{Final decision}\\[1pt]\stepsub{+ confidence (1--7)}};
    \draw[arr] (v) -- (ip);  \draw[arr] (ip) -- (rec);  \draw[arr] (rec) -- (fp);
    \node[font=\small\bfseries, text=cAnthro!75!black, anchor=south west]
      at ($(v.north west)+(0,0.42)$) {Each trial ($\times\,8$)};
    \draw[arr, dashed, draw=cAnthro!55] (st.south) -- ++(0,-0.55) -| (v.north);
  \end{tikzpicture}
  \caption{Study procedure. Top row: the session-level stages, from demographics through
    a practice trial, the main study (highlighted), and the closing voice ratings. Bottom
    row: the fixed four-step sequence repeated on each of the eight trials - participants
    read a symptom vignette, make an initial diagnosis and rate their confidence, hear the
    AI recommendation (which either confirms or contradicts their pick, green box), and
    then make their final diagnosis and confidence rating. The eight trials are split into
    two counterbalanced voice blocks of four, each containing two confirming and two
    contradicting recommendations. The initial-vs.-final decision difference is the bias
    effect analysed throughout.}
  \Description{Flow diagram with two rows of boxes. Top row: demographics, practice, main study, closing. Bottom row: the four per-trial steps from symptom vignette through initial decision, AI recommendation, and final decision.}
  \label{fig:procedure}
\end{figure*}
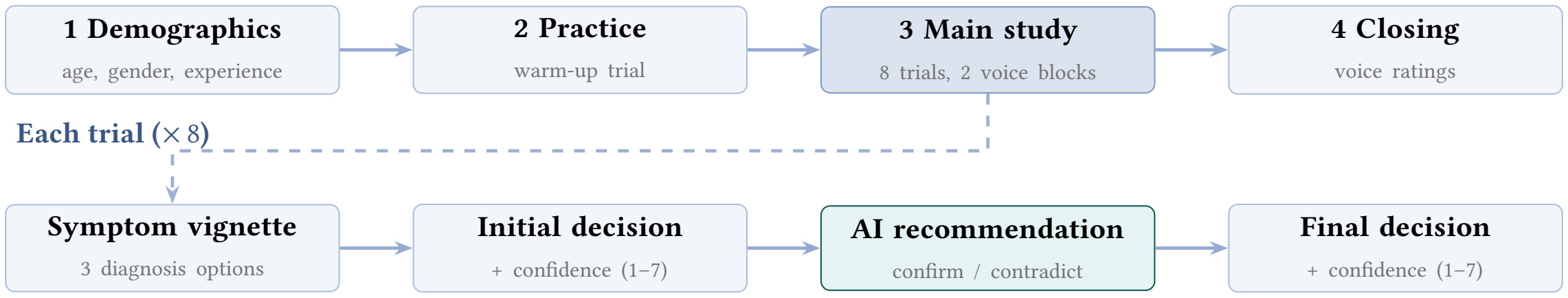

\subsection{Experimental Design}
We conducted a within-subjects online experiment crossing two factors: \emph{AI condition}
(whether the AI recommendation \emph{confirmed} or \emph{contradicted} the participant's
initial diagnosis) and \emph{voice design} (whether it was delivered in an
\emph{anthropomorphic} (i.e., human-like) or a \emph{machine-like} synthetic voice). Every
participant experienced all four combinations of the two factors.

Three dependent variables were measured on each trial: whether the participant
\emph{changed} their diagnosis after hearing the recommendation, the \emph{shift} in their
stated confidence, and the \emph{response time} needed to reach the final decision.
Figure~\ref{fig:procedure} summarises the procedure.

We tested two hypothesis sets: the first, a general confirmation bias toward the AI across
the three outcomes:
\begin{itemize}
  \item[\textbf{H1a}] Participants change their diagnosis more often when the AI contradicts
    their initial choice than when it confirms it.
  \item[\textbf{H1b}] A confirming recommendation increases confidence relative to a
    contradicting one, i.e.\ the confidence shift is more positive after confirmation.
  \item[\textbf{H1c}] Participants take longer to reach their final decision after a
    contradicting recommendation than after a confirming one.
\end{itemize}

The second, primary hypothesis concerns whether the AI's voice moderates this bias, and in
which direction:
\begin{itemize}
  \item[\textbf{H2}] The strength of the confirmation-bias effect (H1a--H1c) differs between
    the anthropomorphic and the machine-like voice.
  \item[\textbf{H2a}] The confirmation-bias effect is stronger for the anthropomorphic voice
    than for the machine-like voice, as human-like delivery is expected to raise the perceived
    credibility of the recommendation.
\end{itemize}

\subsection{Participants}
The final sample consisted of $\sampleN$ participants who completed the study ($\nTrials$
analysed trials); sixteen were recruited during a live event and received a slice of pizza
as compensation, the remainder through convenience sampling within the researchers' extended
personal networks. Five sessions that were started but not completed were excluded; the
sample's demographic composition is reported with the results (see \emph{Sample} at the
start of the Results).

\subsection{Materials and Apparatus}

\begin{figure*}
  \centering
  \begin{minipage}[t]{0.24\textwidth}
    \centering
    \includegraphics[width=\linewidth]{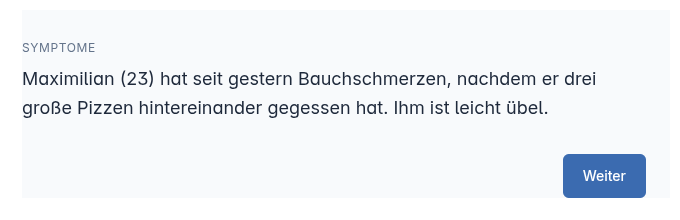}\par
    {\footnotesize (a) Symptom vignette}
  \end{minipage}\hfill
  \begin{minipage}[t]{0.24\textwidth}
    \centering
    \includegraphics[width=\linewidth]{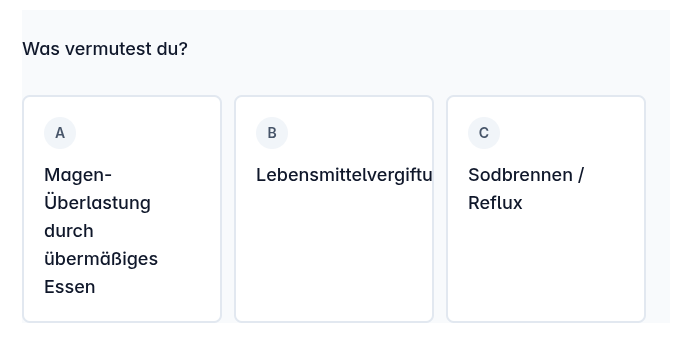}\par
    {\footnotesize (b) Diagnosis choice}
  \end{minipage}\hfill
  \begin{minipage}[t]{0.24\textwidth}
    \centering
    \includegraphics[width=\linewidth]{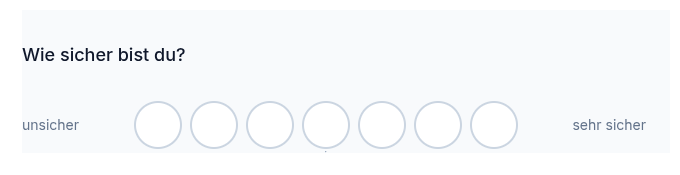}\par
    {\footnotesize (c) Confidence rating}
  \end{minipage}\hfill
  \begin{minipage}[t]{0.24\textwidth}
    \centering
    \includegraphics[width=\linewidth]{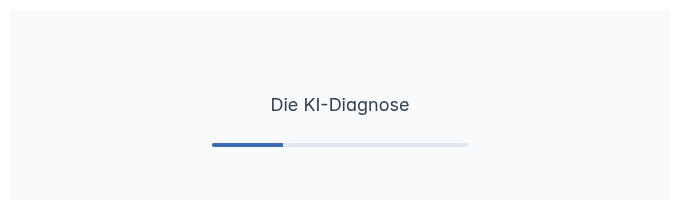}\par
    {\footnotesize (d) AI voice recommendation}
  \end{minipage}
  \caption{The participant-facing interface of the study application (in German, as
    presented to participants), illustrating the four key screens of a trial:
    (a)~the symptom vignette describing the patient case; (b)~the initial diagnosis
    choice among three options; (c)~the seven-point confidence rating (anchored
    \emph{unsicher}/unsure to \emph{sehr sicher}/very sure); and (d)~the spoken AI
    recommendation, presented as audio playback without on-screen text. After the
    recommendation, participants repeated screens (b) and (c) for their final
    decision.}
  \Description{Four screenshots of the German study interface: the symptom vignette text, the three diagnosis option cards, the seven-point confidence rating scale, and the audio playback screen for the AI recommendation.}
  \label{fig:ui}
\end{figure*}

The diagnostic cases were short patient vignettes constructed by the research team from
the disease--symptom associations of a publicly available, artificially generated dataset covering
773 diseases and 377 symptoms~\citep{dhivyeshrk2023}: each vignette describes the symptoms the dataset lists
for one disease, written by the authors in everyday language so that laypeople could reason
about them without specialist knowledge. Each vignette was followed by three diagnostic
options: the disease the symptoms were drawn from and two alternative diagnoses that the
authors judged plausible for the described symptoms. As the authors are not medical
professionals, the cases were not clinically validated. Instead, the vignettes and options
were piloted with peers to ensure that they were understandable and that none of the options
appeared implausible. The study therefore does not rely on the medical correctness of any
option: the manipulation concerns whether the AI confirms or contradicts the participant's
\emph{own} initial choice, and no option was scored as correct in the analysis. The pool
comprised twelve vignettes; to broaden the stimulus set without lengthening the sessions,
each participant received a seeded random subset of eight, and case was modelled as a random
effect in all analyses. The participant-facing interface is shown in Figure~\ref{fig:ui}.

AI recommendations were pre-scripted by the authors rather than produced by a live model, so
that the verbal content was held constant across participants and only the voice varied.
For every option of every case, the authors wrote a short German explanation of two to three
sentences that named the diagnosis and related it to the described symptoms (e.g., that the
symptoms began directly after a large meal, which is typical of the recommended diagnosis);
these texts were then synthesised into speech with the two voices described below. For each option the spoken content was identical across the two
voices; the voices differed only in their acoustic delivery.

The anthropomorphic voice was synthesised with ElevenLabs' neural text-to-speech (TTS)
service
(model \texttt{eleven\_v3}, a custom female voice created with ElevenLabs Voice Design), configured for expressive delivery
(stability $0.35$, similarity boost $0.75$, style $0.45$, speaker boost enabled) and enriched
with natural delivery cues: audible breaths and hesitation fillers (e.g.\ ``ähm'').
The machine-like voice was synthesised with Microsoft Edge TTS (voice
\texttt{de-DE-KatjaNeural}), rendered with a lowered pitch ($-30$\,Hz) and slightly reduced
rate ($-5\%$) to emphasise a flat, plainly synthetic character, and read the same content
without any delivery cues.

The two voices were thus intended to contrast human-like prosody and warmth against flat,
mechanical delivery. To ensure this was a contrast of \emph{character} rather than
\emph{level}, all audio was loudness-normalised to $-16$\,LUFS (EBU R128) with ffmpeg
(\texttt{loudnorm}, $I=-16$, TP$=-1.5$, LRA$=11$).

The study ran in the browser on both desktop and smartphone, and participants were advised
to use headphones for consistent audio playback.

\subsection{Procedure}
Before the main experiment, participants provided demographic information (age and gender)
and reported their prior experience with AI symptom checkers and AI voice assistants. A
single practice trial with one additional vignette then familiarised them with the task and
was excluded from analysis.

For each case, participants first read the patient vignette, selected the diagnosis they
considered most likely, and rated their confidence on a 7-point Likert scale from 1 (very low) to 7 (very
high). They then listened to the AI voice recommendation, after which they again selected one
of the three options and re-rated their confidence. Response times were recorded both before
and after the recommendation.

Each participant's eight cases were split into two blocks of four, one block per voice, with a short break
between the blocks. Each participant thus heard one voice throughout the first block and the
other throughout the second, with no mixing of voices within a block.

Trial assignment was generated deterministically from a per-session seed, which makes each
session fully reproducible and exactly counterbalanced. A seeded pseudo-random generator
(Mulberry32, seeded via an FNV-1a hash of the session identifier) drove all randomisation.
The participant's eight cases were sorted into four pairs, each pair's coin flip assigning
its two cases to opposite AI conditions - yielding exactly four confirming and four
contradicting trials per participant. Block order (anthropomorphic-first vs.\
machine-first) was likewise coin-flip balanced; each block received two confirming and two
contradicting cases, shuffled within the block to reduce presentation-order effects.

After completing all trials, participants heard a short sample of each voice again and
rated it on eight 7-point semantic differentials, which served as a manipulation check.
Human-likeness (machine-like--human-like), naturalness (artificial--natural), pleasantness
(unpleasant--pleasant), and competence (incompetent--competent) were single items adapted from
the anthropomorphism, likeability, and perceived-intelligence scales of the Godspeed
questionnaire~\citep{bartneck2009}, which was developed for robots and uses 5-point scales;
the naturalness item combines its fake--natural and artificial--lifelike pairs. Trustworthiness
and intelligibility followed subjective rating studies of synthetic
voices~\citep{kuehne2020}, warmth (distant--warm) was added following the warmth--competence
framework of social perception~\citep{fiske2007}, and a final item captured whether the voice
sounded male or female. All items were presented in German translations by the authors, in
randomised order. As this set is an ad-hoc adaptation rather than a validated scale, we
analysed the items individually.

\subsection{Measures and Analysis}
The three dependent variables were operationalised from the per-trial records. Diagnosis
\emph{change} was a binary indicator that the final diagnostic choice differed from the
initial one. Confidence \emph{shift} was the difference between the post- and pre-recommendation
confidence ratings on the 1--7 scale. Final-decision \emph{response time} was the time,
recorded in milliseconds and reported in seconds, from the moment the answer options
reappeared - i.e., after the recommendation had finished playing and the participant had
continued - to the final diagnostic choice. Because the study ran unsupervised in the
browser, these times also include any moments in which a participant was not attending to
the task (see Limitations).

We analysed the trial-level data ($\nTrials$ trials) with one mixed-effects model per
outcome, each with fixed effects for AI condition, voice, and their interaction and crossed
random intercepts for participant and case. Both factors were coded with sum-to-zero
contrasts (confirming AI condition and machine-like voice as the positively signed levels),
so that each main effect is averaged over the levels of the other factor. Diagnosis change
was modelled with a binomial generalised linear mixed model (GLMM, logit
link)~\citep{bates2015}. For the two continuous outcomes, the residuals of Gaussian linear
mixed models (LMMs) deviated clearly from normality, with heavy tails for the confidence shift
and a pronounced right skew for response time even after log transformation (Shapiro--Wilk
$W=\shapiroShiftW$ and $W=\shapiroRtW$, both $p\shapiroShiftP$; Q--Q plots in
Appendix~\ref{app:residuals}). We therefore modelled the confidence shift with a robust LMM,
which down-weights outlying observations~\citep{koller2016}, and response time with a Gamma
GLMM with log link on untransformed seconds~\citep{lo2015,brooks2017}, which accommodates the
right-skewed distribution of response times without transforming the outcome. For the
confidence shift, the variance of the case intercept was estimated at zero (singular fit),
i.e.\ the shift did not vary systematically between cases; this leaves the fixed effects
unaffected.

From each model we derived estimated marginal means for the four AI\,$\times$\,voice cells
and the contrasts of interest on the model's natural scale: the odds ratio of a diagnosis
change (contradict vs.\ confirm), the difference in confidence shift (confirm $-$
contradict, in scale points), and the ratio of mean response times (contradict / confirm).
Voice moderation (H2) was tested by the AI\,$\times$\,voice interaction, expressed as the
difference between the AI effects under the two voices. All tests and 95\,\% confidence
intervals (CIs) are Wald-type. As sensitivity checks, we refit the two continuous outcomes
with the Gaussian LMMs used in an earlier version of this analysis
(Appendix~\ref{app:sensitivity}) and ran participant-level nonparametric tests on aggregated
data (Appendix~\ref{app:participant}).

\subsection{Ethical Considerations}
Participation was voluntary and anonymous: no clinically identifying information was
collected, and participants could abort at any time without giving a reason. Because the AI
recommendations were simulated and deliberately contradicted participants on half of the
trials, every participant received a debriefing explaining that the recommendations had been
pre-configured and did not constitute medical advice; the use of vignettes adapted for
laypeople, rather than participants' own health data, further limited any risk.

\section{Results}

This section first describes the sample and then reports our two sets of hypotheses: a
general confirmation bias toward the AI (H1a--H1c) and the moderating role of voice design
(H2).

\subsection{Sample}
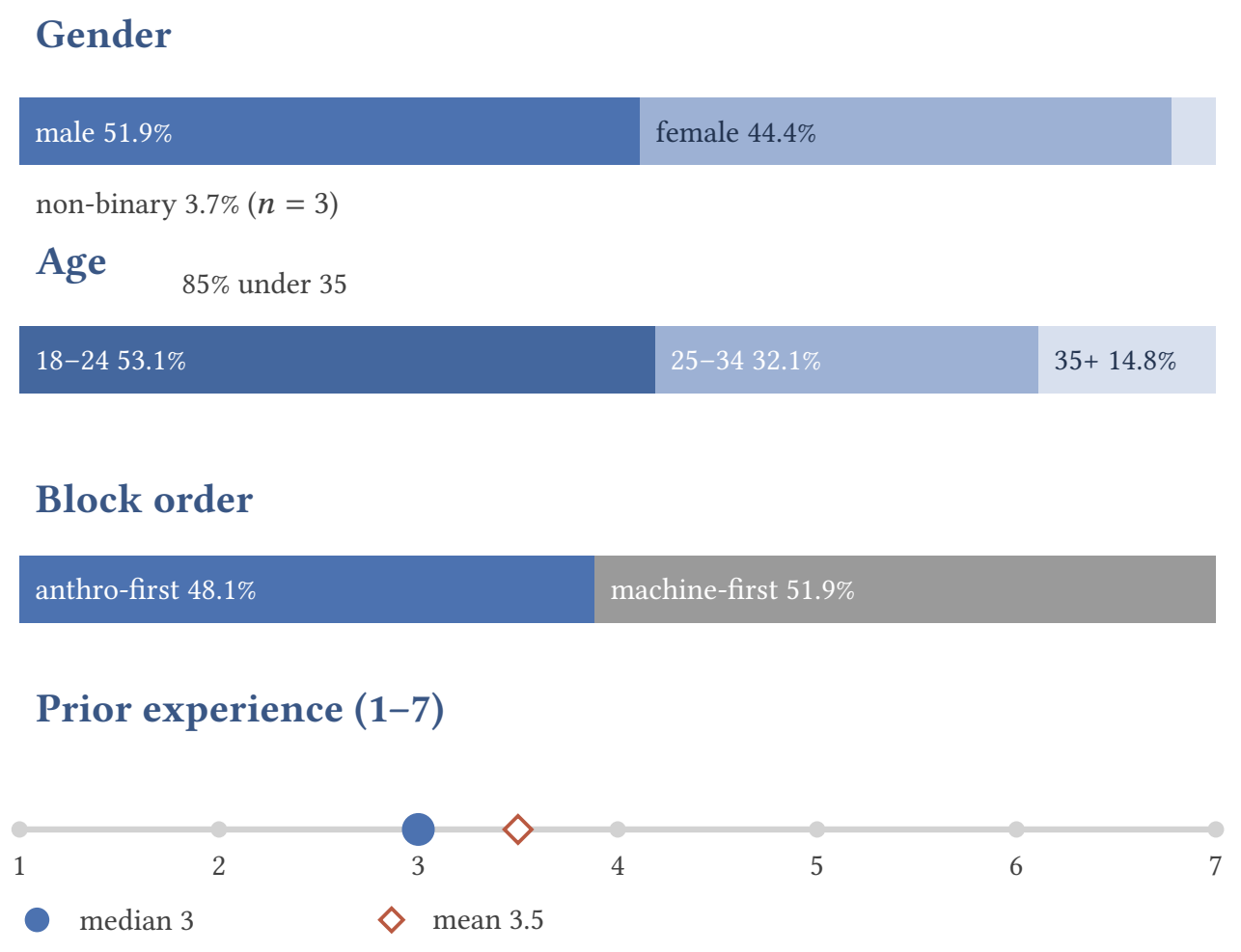
\begin{figure}
  \centering
  \begin{tikzpicture}[
      font=\small,
      cat/.style   ={font=\small\bfseries, text=cAnthro!75!black, anchor=south west},
      inlbl/.style ={font=\scriptsize, anchor=west, inner sep=0pt},
      note/.style  ={font=\scriptsize, text=black!75, anchor=west},
      tk/.style    ={font=\scriptsize, text=black!75},
    ]
    \def\W{8.2}\def\H{0.46}
    \node[cat] at (0,2pt) {Gender};
    \begin{scope}[shift={(0,-0.62)}]
      \pgfmathsetmacro\xa{\demoMalePct/100*\W}
      \pgfmathsetmacro\xb{(\demoMalePct+\demoFemPct)/100*\W}
      \fill[cAnthro]    (0,0)   rectangle (\xa,\H);
      \fill[cAnthro!55] (\xa,0) rectangle (\xb,\H);
      \fill[cAnthro!22] (\xb,0) rectangle (\W,\H);
      \node[inlbl, text=white]          at (0.1,\H/2)     {male \demoMalePct\%};
      \node[inlbl, text=cAnthro!45!black] at (\xa+0.1,\H/2) {female \demoFemPct\%};
      \node[note, anchor=north west] at (0,-0.07) {non-binary \demoNbPct\% ($n=\demoNbN$)};
    \end{scope}
    \node[cat] at (0,-1.55) {Age};
    \node[note] at (1.0,-1.43) {\demoUnderThirtyfive\% under 35};
    \begin{scope}[shift={(0,-2.19)}]
      \pgfmathsetmacro\ya{\demoAgeYoungPct/100*\W}
      \pgfmathsetmacro\yb{(\demoAgeYoungPct+\demoAgeMidPct)/100*\W}
      \fill[cAnthro!90!black] (0,0)   rectangle (\ya,\H);
      \fill[cAnthro!55]       (\ya,0) rectangle (\yb,\H);
      \fill[cAnthro!22]       (\yb,0) rectangle (\W,\H);
      \node[inlbl, text=white]            at (0.1,\H/2)    {18--24 \demoAgeYoungPct\%};
      \node[inlbl, text=white]            at (\ya+0.1,\H/2){25--34 \demoAgeMidPct\%};
      \node[inlbl, text=cAnthro!45!black] at (\yb+0.1,\H/2){35+ \demoAgeOldPct\%};
    \end{scope}
    \node[cat] at (0,-3.12) {Block order};
    \begin{scope}[shift={(0,-3.76)}]
      \pgfmathsetmacro\ba{\demoAnthroFirstPct/100*\W}
      \fill[cAnthro]  (0,0)   rectangle (\ba,\H);
      \fill[cMachine] (\ba,0) rectangle (\W,\H);
      \node[inlbl, text=white] at (0.1,\H/2)    {anthro-first \demoAnthroFirstPct\%};
      \node[inlbl, text=white] at (\ba+0.1,\H/2){machine-first \demoMachineFirstPct\%};
    \end{scope}
    \node[cat] at (0,-4.62) {Prior experience (1--7)};
    \begin{scope}[shift={(0,-5.18)}]
      \draw[gray!35, line width=1.2pt] (0,0) -- (\W,0);
      \foreach \i in {1,...,7}{
        \pgfmathsetmacro\xx{(\i-1)/6*\W}
        \fill[gray!35] (\xx,0) circle (1.6pt);
        \node[tk, below=2pt] at (\xx,0) {\i};
      }
      \pgfmathsetmacro\xmed{(\expScMed-1)/6*\W}
      \pgfmathsetmacro\xmean{(\expMean-1)/6*\W}
      \fill[cAnthro] (\xmed,0) circle (3.2pt);
      \node[diamond, draw=cRefute!80!black, fill=white, line width=0.7pt, inner sep=1.4pt] at (\xmean,0) {};
      \fill[cAnthro] (0.12,-0.62) circle (2.4pt);
      \node[anchor=west, font=\scriptsize, text=black!75] at (0.30,-0.62) {median \expScMed};
      \node[diamond, draw=cRefute!80!black, fill=white, line width=0.7pt, inner sep=1.2pt] at (2.55,-0.62) {};
      \node[anchor=west, font=\scriptsize, text=black!75] at (2.72,-0.62) {mean \expMean};
    \end{scope}
  \end{tikzpicture}
  \caption{Sample overview ($N=\sampleN$). The stacked bar chart shows the composition
    of the sample by gender, age band, and counterbalanced block order (each bar sums
    to 100\,\%, with the within-segment percentages printed inline); the bottom row
    places the median (filled dot) and mean (open diamond) of prior experience with AI
    symptom checkers and voice assistants on the 1--7 scale.}
  \Description{Stacked bars showing sample composition by gender, age, and block order, plus a 1-to-7 line marking median and mean prior experience near the lower end of the scale.}
  \label{fig:sample}
\end{figure}

All $\sampleN$ participants who completed the study were included (Figure~\ref{fig:sample}).
The sample was roughly balanced by gender and skewed young (\demoUnderThirtyfive\% under
35). The two block orders were represented in similar numbers: \demoAnthroFirstN{}
participants heard the anthropomorphic voice first and \demoMachineFirstN{} the machine-like
voice first. Participants reported little prior experience with AI symptom checkers or voice
AI assistants, with both medians at the lower end of the 1--7 scale (median \expScMed).

\subsection{General Confirmation Bias (H1)}
Full model coefficients are reported in Table~\ref{tab:coef} and model-based estimates for
each condition in Figure~\ref{fig:h1results}.

\begin{table}
  \centering
  \caption{Fixed-effect coefficients of the three primary models
    (\(N=81\) participants, 648 trials; crossed random intercepts for participant and case).
    Factors are sum-to-zero coded: \emph{AI} \(=\) confirm (\(+1\)) vs.\ contradict (\(-1\)),
    \emph{Voice} \(=\) machine-like (\(+1\)) vs.\ anthropomorphic (\(-1\)), so each
    coefficient is half the difference between the two levels on the model's link scale.
    All tests and 95\,\% CIs are Wald-type.}
  \label{tab:coef}
  \small
  \setlength{\tabcolsep}{3.5pt}
  \begin{tabular}{lrrrrl}
    \toprule
    \multicolumn{6}{l}{\textbf{(A) Diagnosis change} (binomial GLMM, logit link)}\\
    Term & \(b\) & SE & \(z\) & \(p\) & \(e^{b}\) [95\%~CI]\\
    \midrule
    Intercept & $-2.976$ & $0.395$ & $-7.54$ & ${<}.001$ & $0.05$ [$0.02$, $0.11$]\\
    AI & $-2.719$ & $0.338$ & $-8.04$ & ${<}.001$ & $0.07$ [$0.03$, $0.13$]\\
    Voice & $0.249$ & $0.302$ & $0.82$ & $.410$ & $1.28$ [$0.71$, $2.32$]\\
    AI\,$\times$\,Voice & $0.325$ & $0.302$ & $1.08$ & $.282$ & $1.38$ [$0.77$, $2.50$]\\
    \bottomrule
  \end{tabular}

  \vspace{4pt}
  \begin{tabular}{lrrrrl}
    \toprule
    \multicolumn{6}{l}{\textbf{(B) Confidence shift} (robust LMM)}\\
    Term & \(b\) & SE & \(t\) & \(p\) & 95\%~CI\\
    \midrule
    Intercept & $0.314$ & $0.055$ & $5.75$ & ${<}.001$ & [$0.207$, $0.421$]\\
    AI & $0.505$ & $0.037$ & $13.79$ & ${<}.001$ & [$0.434$, $0.577$]\\
    Voice & $0.003$ & $0.037$ & $0.09$ & $.928$ & [$-0.069$, $0.075$]\\
    AI\,$\times$\,Voice & $0.011$ & $0.037$ & $0.31$ & $.759$ & [$-0.061$, $0.083$]\\
    \bottomrule
  \end{tabular}

  \vspace{4pt}
  \begin{tabular}{lrrrrl}
    \toprule
    \multicolumn{6}{l}{\textbf{(C) Response time} (Gamma GLMM, log link, seconds)}\\
    Term & \(b\) & SE & \(z\) & \(p\) & \(e^{b}\) [95\%~CI]\\
    \midrule
    Intercept & $1.248$ & $0.062$ & $20.19$ & ${<}.001$ & $3.48$ [$3.09$, $3.93$]\\
    AI & $-0.299$ & $0.026$ & $-11.34$ & ${<}.001$ & $0.74$ [$0.70$, $0.78$]\\
    Voice & $0.016$ & $0.026$ & $0.62$ & $.533$ & $1.02$ [$0.97$, $1.07$]\\
    AI\,$\times$\,Voice & $-0.021$ & $0.026$ & $-0.81$ & $.420$ & $0.98$ [$0.93$, $1.03$]\\
    \bottomrule
  \end{tabular}
\end{table}

\pgfplotstableread{generated/fig-pick-change.dat}\pickdat
\pgfplotstableread{generated/fig-confidence.dat}\confshift
\pgfplotstableread{generated/fig-rt.dat}\rtdat
\begin{figure*}
  \centering
  \begin{tikzpicture}
    \begin{groupplot}[
        group style={group size=3 by 1, horizontal sep=1.5cm},
        studbase, width=5.55cm, height=4.4cm,
        title style={font=\small},
      ]
      \nextgroupplot[
        bar width=8pt,
        title={(a) Diagnosis changed},
        ylabel={Diagnosis changed (\%)},
        symbolic x coords={anthro,machine}, xtick=data,
        xticklabels={Anthro., Machine},
        ymin=0, ymax=72, ytick={0,20,40,60},
      ]
        \addplot[draw=cConfirm!60!black, fill=cConfirm, bar shift=-5.5pt, studerr]
          table[x=voice, y=confirm, y error plus=confirm_ehi, y error minus=confirm_elo]{\pickdat};
        \addplot[draw=cRefute!60!black, fill=cRefute, bar shift=5.5pt, studerr]
          table[x=voice, y=refute, y error plus=refute_ehi, y error minus=refute_elo]{\pickdat};
        \addplot[studval, bar shift=-5.5pt,
          nodes near coords={\pgfmathprintnumber[fixed,precision=1]{\pgfplotspointmeta}\%}]
          table[x=voice, y expr=\thisrow{confirm}+\thisrow{confirm_ehi}, meta=confirm]{\pickdat};
        \addplot[studval, bar shift=5.5pt,
          nodes near coords={\pgfmathprintnumber[fixed,precision=1]{\pgfplotspointmeta}\%}]
          table[x=voice, y expr=\thisrow{refute}+\thisrow{refute_ehi}, meta=refute]{\pickdat};

      \nextgroupplot[
        bar width=8pt,
        title={(b) Confidence shift},
        ylabel={Confidence shift (post\,$-$\,pre)},
        symbolic x coords={anthro,machine}, xtick=data,
        xticklabels={Anthro., Machine},
        ymin=-0.5, ymax=1.5, ytick={-0.5,0,0.5,1,1.5},
      ]
        \addplot[draw=cConfirm!60!black, fill=cConfirm, bar shift=-5.5pt, studerr]
          table[x=voice, y=confirm, y error plus=confirm_ehi, y error minus=confirm_elo]{\confshift};
        \addplot[draw=cRefute!60!black, fill=cRefute, bar shift=5.5pt, studerr]
          table[x=voice, y=refute, y error plus=refute_ehi, y error minus=refute_elo]{\confshift};
        \addplot[studval, bar shift=-5.5pt,
          nodes near coords={\pgfmathprintnumber[fixed,fixed zerofill,precision=2]{\pgfplotspointmeta}}]
          table[x=voice, y expr=\thisrow{confirm}+\thisrow{confirm_ehi}, meta=confirm]{\confshift};
        \addplot[studval, bar shift=5.5pt,
          nodes near coords={\pgfmathprintnumber[fixed,fixed zerofill,precision=2]{\pgfplotspointmeta}}]
          table[x=voice, y expr=\thisrow{refute}+\thisrow{refute_ehi}, meta=refute]{\confshift};

      \nextgroupplot[
        bar width=8pt,
        title={(c) Response time},
        ylabel={Final-decision time (s)},
        symbolic x coords={anthro,machine}, xtick=data,
        xticklabels={Anthro., Machine},
        ymin=0, ymax=6.9, ytick={0,1,2,3,4,5,6},
      ]
        \addplot[draw=cConfirm!60!black, fill=cConfirm, bar shift=-5.5pt, studerr]
          table[x=voice, y=confirm, y error plus=confirm_ehi, y error minus=confirm_elo]{\rtdat};
        \addplot[draw=cRefute!60!black, fill=cRefute, bar shift=5.5pt, studerr]
          table[x=voice, y=refute, y error plus=refute_ehi, y error minus=refute_elo]{\rtdat};
        \addplot[studval, bar shift=-5.5pt,
          nodes near coords={\pgfmathprintnumber[fixed,fixed zerofill,precision=2]{\pgfplotspointmeta}}]
          table[x=voice, y expr=\thisrow{confirm}+\thisrow{confirm_ehi}, meta=confirm]{\rtdat};
        \addplot[studval, bar shift=5.5pt,
          nodes near coords={\pgfmathprintnumber[fixed,fixed zerofill,precision=2]{\pgfplotspointmeta}}]
          table[x=voice, y expr=\thisrow{refute}+\thisrow{refute_ehi}, meta=refute]{\rtdat};
    \end{groupplot}
    \node[anchor=north, font=\small, inner sep=0pt] at ($(group c2r1.south)+(0,-0.75cm)$) {%
      \tikz[baseline=-0.6ex]\fill[draw=cConfirm!60!black, fill=cConfirm] (0,-0.1) rectangle (0.3,0.1);~AI confirms\hspace{14pt}%
      \tikz[baseline=-0.6ex]\fill[draw=cRefute!60!black, fill=cRefute] (0,-0.1) rectangle (0.3,0.1);~AI contradicts};
  \end{tikzpicture}
  \caption{The three trial-level outcomes tested under H1, by voice (x-axis within each
    panel) and recommendation type (teal = AI confirms, red = AI contradicts; see legend): (a)
    model-estimated rate of trials on which the initial diagnosis was revised; (b) change in self-reported confidence
    (post\,$-$\,pre, 1--7 scale); and (c) time to the final decision in seconds.
    Bars are estimated marginal means from the binomial GLMM (a), the robust LMM (b), and the
    Gamma GLMM (c), with 95\,\% CI whiskers; the number above each bar is its point estimate. Model
    coefficients are given in Table~\ref{tab:coef}.}
  \Description{Three grouped bar charts side by side. (a) Probability of diagnosis change is
    near zero when the AI confirms and roughly 40 to 45 percent when it contradicts, for both
    voices. (b) Confidence rises by about one scale point when the AI confirms and stays near
    zero when it contradicts, for both voices. (c) Final decisions take about 2.2 seconds
    when the AI confirms and about 3.6 seconds when it contradicts, for both voices.}
  \label{fig:h1results}
\end{figure*}
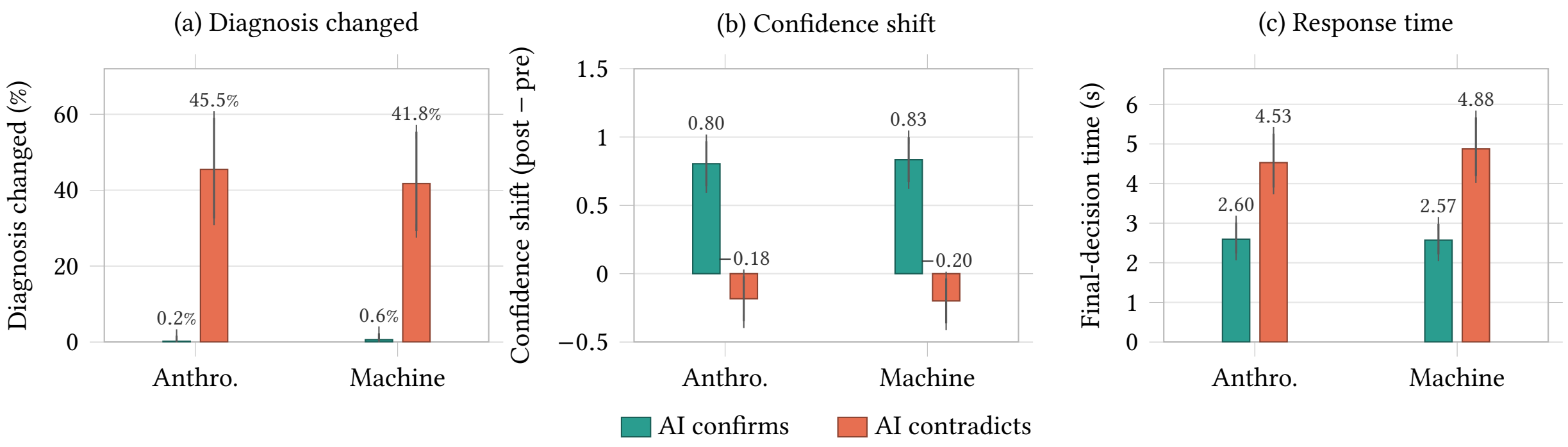

\paragraph{Diagnostic choice (H1a).} The AI recommendation strongly influenced the final
diagnosis. Participants changed their diagnosis in \changeRefute{} of trials when the AI
contradicted them, but in only \changeConfirm{} when it confirmed them (change rate = trials
with a changed diagnosis divided by all trials in that condition;
Figure~\ref{fig:h1results}a). In the binomial GLMM, the odds of a change were about
\pickOR{} times higher after a contradicting than after a confirming recommendation
($\text{OR}=\pickOR$, 95\,\% CI [\pickORlo, \pickORhi], $p\pickAiP$; Table~\ref{tab:coef},
panel A). The very wide interval reflects the near-absence of changes under confirmation (see
Limitations); the direction and significance of the effect are unaffected. H1a is supported.

\paragraph{Confidence (H1b).} A confirming recommendation raised confidence substantially
(from \confPriorConfirm{} to \confAfterConfirm{} on the 1--7 scale, a shift of
$+\shiftConfirm$), whereas a contradicting one lowered it only slightly (from
\confPriorRefute{} to \confAfterRefute{}, a shift of $\shiftRefute$). In the robust LMM, the
confidence shift was \shiftEff{} scale points more positive after confirmation than after
contradiction (95\,\% CI [\shiftEffLo, \shiftEffHi], $p\shiftAiP$;
Figure~\ref{fig:h1results}b; Table~\ref{tab:coef}, panel B). H1b is supported. Because
confidence after a switch refers to a different diagnosis than before, we repeated the
analysis on the \stayerPct{}\,\% of trials on which participants kept their initial pick
(robust LMM fitted to these trials only); the difference remained significant and comparable
in size ($\Delta=\stayerShiftDiff$, $p\stayerShiftP$).

\paragraph{Response time (H1c).} Decisions took markedly longer after a contradicting
recommendation. Raw mean decision times were \rtConfirm{}\,s under confirmation versus
\rtRefute{}\,s under contradiction. The Gamma GLMM estimated mean times of
\rtConfirmModel{}\,s and \rtRefuteModel{}\,s, respectively, i.e.\ final decisions took
\rtRatio{} times as long after contradiction (95\,\% CI [\rtRatioLo, \rtRatioHi],
$p\rtAiP$; Figure~\ref{fig:h1results}c; Table~\ref{tab:coef}, panel C), or, equivalently,
were \rtFasterPct{} faster after confirmation. H1c is supported.

The participant-level nonparametric checks in Appendix~\ref{app:participant} confirm all
three effects.

\subsection{Influence of Voice Design (H2)}
\label{sec:h2}
Against our primary hypothesis, the anthropomorphic and machine-like voices produced
indistinguishable effects on every outcome. Neither the voice main effect nor the
AI\,$\times$\,voice interaction was significant in any of the three models (diagnosis change:
voice $p\pickVoiceP$, interaction $p\pickIntP$; confidence shift: voice $p\shiftVoiceP$,
interaction $p\shiftIntP$; response time: voice $p\rtVoiceP$, interaction $p\rtIntP$).
Table~\ref{tab:voiceeffects} shows the AI effect estimated separately within each voice: it
was large, significant, and of similar size for both voices on all three outcomes. For the
confidence shift, the difference between the voices' AI effects was \shiftInt{} scale points
(95\,\% CI [\shiftIntLo, \shiftIntHi]); for response time, the ratio of the two voices' AI
effects was \rtIntRatio{} (95\,\% CI [\rtIntRatioLo, \rtIntRatioHi]). Both intervals include
no difference and are small relative to the AI effect itself (\shiftEff{} points and a ratio
of \rtRatio{}, respectively), so a moderation of practically relevant size is unlikely for
these outcomes. For diagnosis change, the corresponding ratio of odds ratios was
\pickIntRatio{} (95\,\% CI [\pickIntRatioLo, \pickIntRatioHi]); this interval is wide because
of the sparse confirm-condition cell. H2 is not supported, and consequently neither is the
directional prediction H2a.

\begin{table}
  \centering
  \caption{Effect of the AI condition within each voice and the difference between voices
    (interaction), from the primary models with 95\,\% CIs. Diagnosis change: odds ratio
    contradict/confirm; confidence shift: confirm \(-\) contradict in scale points; response
    time: ratio of mean times contradict/confirm. For the interaction row, the voice
    difference is expressed as a ratio of the anthropomorphic to the machine-like effect for
    odds and time, and as anthropomorphic \(-\) machine-like for confidence; no difference
    corresponds to 1 and 0, respectively.}
  \label{tab:voiceeffects}
  \small
  \setlength{\tabcolsep}{3pt}
  \begin{tabular}{llll}
    \toprule
     & Change (OR) & Confidence (\(\Delta\)) & Time (ratio)\\
    \midrule
    Anthropomorphic & $440$ [$53$, $3640$] & $0.99$ [$0.79$, $1.19$] & $1.74$ [$1.51$, $2.01$]\\
    Machine-like & $120$ [$31$, $467$] & $1.03$ [$0.83$, $1.24$] & $1.90$ [$1.64$, $2.19$]\\
    \midrule
    Interaction & $3.66$ [$0.34$, $38.99$] & $-0.05$ [$-0.33$, $0.24$] & $0.92$ [$0.75$, $1.13$]\\
    \bottomrule
  \end{tabular}
\end{table}

\subsection{Manipulation Check}
To verify that the two voices were perceived differently, participants rated each voice on
a reduced set of Likert items at the end of the study (Figure~\ref{fig:manipulation}). The
anthropomorphic voice was rated significantly more human-like, natural, warm, pleasant,
intelligible, and trustworthy, and it sounded more clearly female than the machine-like
voice, which was rated near the scale midpoint on the male--female item (all $p<.05$). The
sole exception was perceived competence, which did not differ between the voices
($p=.89$). The manipulation was therefore successful: the null effect of voice on bias
cannot be attributed to participants failing to perceive the voices as different. We note,
however, that both voices were rated as fairly artificial in absolute terms, a point we
return to in the discussion.

\pgfplotstableread{generated/fig-manipulation.dat}\manipdata
\begin{figure}
  \centering
  \begin{tikzpicture}
    \begin{axis}[
      xbar, bar width=6pt,
      width=0.84\columnwidth, height=6.4cm,
      xmin=1, xmax=7.6, xtick={1,3,5,7},
      xlabel={Rating (1--7)},
      xmajorgrids, grid style={gray!20, thin},
      axis line style={gray!55}, tick style={gray!55},
      tick label style={font=\small}, every axis label/.append style={font=\small},
      symbolic y coords={competence,intelligibility,genderedness,trust,pleasantness,warmth,naturalness,humanness},
      ytick={competence,intelligibility,genderedness,trust,pleasantness,warmth,naturalness,humanness},
      yticklabels={competence,intelligibility,female (vs.\ male),trust,pleasantness,warmth,naturalness,human-likeness},
      yticklabel style={font=\small},
      enlarge y limits=0.10,
      nodes near coords, nodes near coords align={horizontal},
      every node near coord/.append style={anchor=west, font=\scriptsize, text=black!80,
        /pgf/number format/fixed, /pgf/number format/precision=1},
      legend cell align=left,
      legend style={draw=none, fill=none, font=\small, at={(0.5,-0.16)}, anchor=north, legend columns=-1,
        /tikz/every even column/.append style={column sep=10pt}},
    ]
      \addplot[draw=cAnthro!60!black, fill=cAnthro]  table[x=anthro,y=item]{\manipdata};
      \addplot[draw=cMachine!50!black, fill=cMachine] table[x=machine,y=item]{\manipdata};
      \legend{Anthropomorphic, Machine-like}
      \node[anchor=west, font=\scriptsize, text=black!80] at (axis cs:5.0,competence) {n.s.};
    \end{axis}
  \end{tikzpicture}
  \caption{Manipulation check: mean ratings of the two voices on eight perception items (1--7 scale; blue = anthropomorphic voice, grey = machine-like voice, value printed at each bar's end; ``n.s.'' marks the only item without a significant difference between the voices).}
  \Description{Horizontal grouped bar chart of eight voice-perception ratings. The anthropomorphic voice scores higher on every item except competence, which is rated equally for both voices.}
  \label{fig:manipulation}
\end{figure}
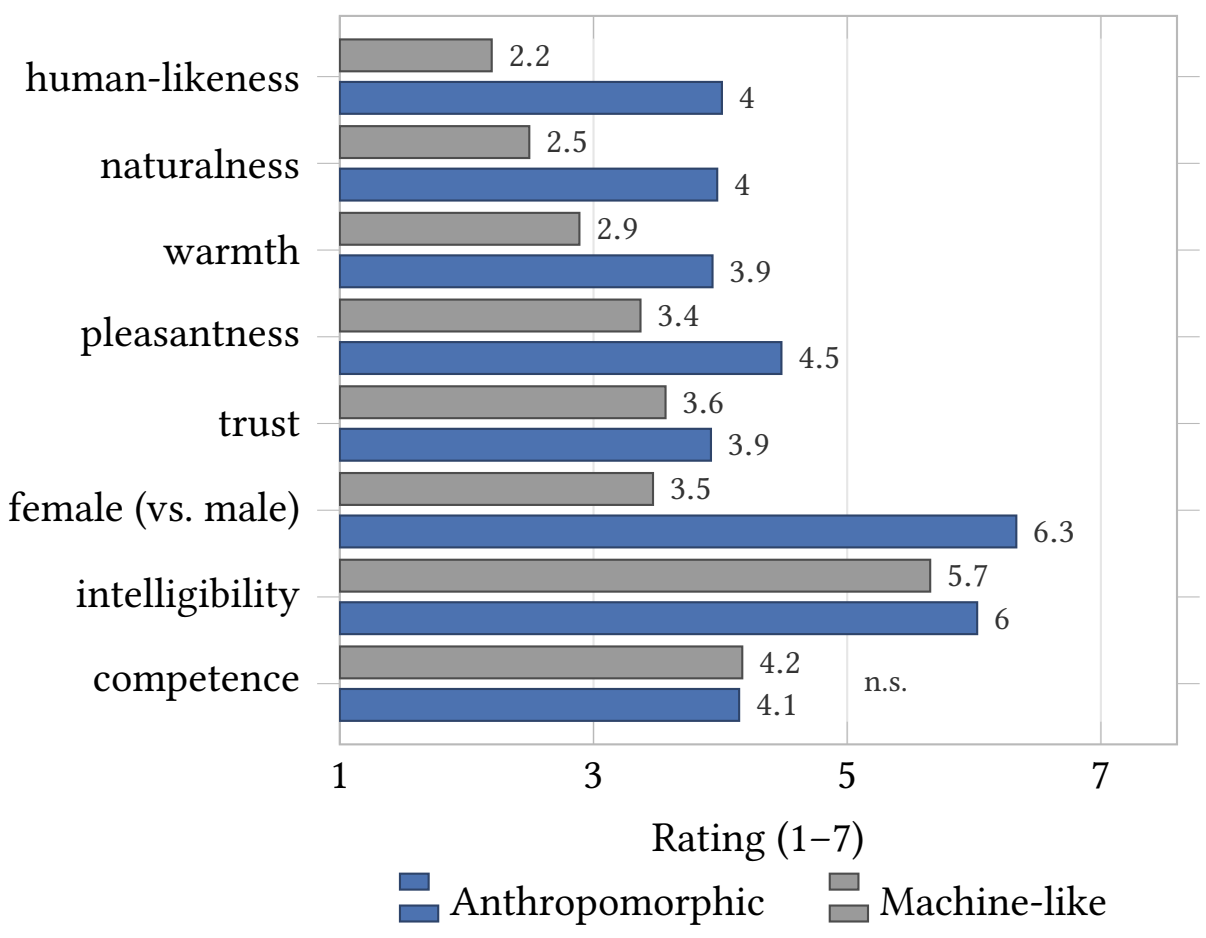

\section{Discussion}

This study asked whether the voice design of an AI symptom checker influences confirmation
bias in AI-assisted self-diagnosis. We found a strong, direction-dependent confirmation
effect: participants were substantially more likely to keep their initial diagnosis when
the AI confirmed it, while contradicting advice produced more frequent changes and longer
response times; confidence, in contrast, rose markedly after confirmation but barely moved
after contradiction. This effect did not differ
between the anthropomorphic and the machine-like voice.

\subsection{A Confirmation Bias, Not a Between-Person Artefact}

Participants did not treat the AI recommendation as neutral information: its direction
shaped both their choices and their subjective confidence. The clearest signature is the
\emph{asymmetric confidence update}: agreement raised confidence
substantially (a shift of $+\shiftConfirm$ scale points) while disagreement barely moved it
($\shiftRefute$), the two differing by $\shiftDiff$ points. This is precisely what
confirmation bias predicts - confirming evidence is weighted more heavily than disconfirming
evidence, and people preferentially accommodate information that agrees with a belief they
already hold~\citep{nickerson1998}. The pattern mirrors advice studies in which a partner's
agreement raised confidence more than disagreement lowered it~\citep{kappes2020} and people
trusted a spurious consensus of agreeing opinions while discounting conflicting
ones~\citep{yaniv2009}; in
perceptual decisions, high confidence has moreover been shown to selectively attenuate the
neural processing of disconfirming post-decision evidence~\citep{rollwage2020}. An updater
equally sensitive to both directions of evidence would lower confidence after contradiction
about as much as it raises it after confirmation, while a sceptical updater would do neither;
our participants raised their confidence after confirmation but barely lowered it after
contradiction.

The small average change after contradiction conceals two opposite movements. Participants
who kept their diagnosis despite the contradiction lowered their confidence
(\refStayShift{} points, \refStayN{} trials), whereas those who switched to the AI's
suggestion rated their confidence in the new diagnosis slightly higher than in their original
one (\refSwitchShift{} points, \refSwitchN{} trials). Contradiction therefore did affect
participants: it either moderately shook their confidence or led them to adopt the AI's view
with about the confidence they had initially placed in their own. In neither case did it
produce a confidence loss comparable to the gain after confirmation.

Crucially, the design rules out the most obvious alternative: because every participant
responded to both confirming and contradicting recommendations, the asymmetry is a
within-person contrast, not an artefact of stable differences in how readily people revise
decisions. A generally reluctant participant would move little in \emph{either} condition,
yet the same people moved far more after contradiction than after confirmation, with the
confidence gap holding within individuals once repeated measures were modelled - a property
of how each person handled agreement versus disagreement, not between-person heterogeneity
in trust or stubbornness.

We are nonetheless careful about the scope of this claim: it characterises the bias at the
level of the sample, not each individual as biased, since we did not separate participants by
their general tendency to trust or distrust the AI (see Limitations). The high rate of
diagnosis changes after contradiction shows that users were also readily willing to abandon
their own judgement when the AI offered an alternative - a strong automation-bias signature
that echoes concerns about over-reliance on AI advice~\citep{kuecking2024,qazi2026}. Taken on
its own, the change rate could therefore be explained by automation bias alone: deferring to
the AI produces many changes after contradiction and none after confirmation, regardless of
any preference for confirming information. The confidence and response-time results go beyond
this. Pure deference would predict comparable updating in both directions, yet confirmation
raised confidence substantially while contradiction barely lowered it, and agreeing advice was
processed markedly faster than disagreeing advice. Taken together, the three outcomes indicate
that confirmation bias, and not only automation bias, shapes how users decide with AI advice.

\subsection{Why Voice Design Did Not Matter}

The absence of a voice effect is the more surprising result, given that our manipulation
check confirmed the two voices were perceived very differently on human-likeness, warmth,
naturalness, and trust. One explanation is that the \emph{content} of the recommendation
simply dominated its delivery, with acoustic style playing at most a secondary role. This
would be in line with findings that advice is used similarly once heeded, whether it is
labelled as coming from an AI or from human peers~\citep{vodrahalli2022}, and that human-like
voices, while generally better liked, have inconsistent effects on behaviour~\citep{seaborn2021}.
A second is that, although the voices differed \emph{relative} to each other, both were rated
as fairly artificial in absolute terms, so the contrast may not have been strong or
human-like enough to change susceptibility to the confirmation effect.

Either way, the null carries a design lesson: prior work has warned that anthropomorphic
design in medical AI may confuse patients about where automated advice ends and human
medical care begins, motivating calls for deliberately machine-like (xenomorphic)
systems~\citep{milford2025}, but our findings show voice style alone did not measurably
change susceptibility to confirmation-related bias, while the recommendation's direction did.
Designers should therefore not treat voice style as a primary lever for reducing biased
decision-making - the leverage lies in what the system says, not in how human it sounds.

\subsection{Implications}

Our findings matter because symptom checkers are typically used outside professional
supervision, where an inaccurate AI can steer laypeople in either direction: a confirming
AI can entrench a wrong self-diagnosis through the very confidence boost we observed, while
a contradicting AI can just as easily talk a correct user out of it, given the high revision
rate we found. In both cases the recommendation's direction, rather than its accuracy,
drives the decision - exactly the failure mode that makes over-reliance on unsupervised
medical AI consequential~\citep{kuecking2024,qazi2026,rosbach2024,cross2024}.

Because the effect was carried by content rather than voice, mitigation should target the
content of the interaction: systems should \emph{communicate uncertainty} by calibrating
expressed confidence to the recommendation's reliability, \emph{prompt users to consider
alternative diagnoses} instead of presenting a single verdict, use \emph{reflective prompts}
that ask users to consider why their initial guess might be wrong, and carry \emph{explicit
reminders that AI advice is not a substitute for professional care}, especially before a
confident-sounding confirmation. Given the null voice effect, these content-level choices
matter more than acoustic or anthropomorphic styling: rather than sounding more (or
deliberately less~\citep{milford2025}) human, patient-facing AI should invest in how it
frames uncertainty and invites users to question their own assumptions.

These implications should be read against the profile of our sample, which skewed young
(\demoUnderThirtyfive\% under 35). Younger, likely more technology-fluent users may relate
to AI advice differently from older populations who make up a large share of real
symptom-checker use, so the size - though not necessarily the existence - of the
confirmation effect may vary across age groups; we discuss this generalisability question
further in the Limitations.

\section{Limitations and Future Work}

Several limitations should be considered when interpreting these results. The change rate
under confirmation was near zero (\confirmChangeN{} of \confirmChangeTotalN{} trials), so the
GLMM's confirm-vs-refute contrast for diagnostic choice is estimated from a sparse cell and
should be read qualitatively rather than as a precise point estimate.

The diagnostic cases were selected and adapted by the researchers, who are not medical
experts, so their clinical accuracy and difficulty were not formally validated - less
critical for our aim, since we tested whether decision and confidence changed after
confirming or contradicting AI advice, not whether participants chose the medically correct
diagnosis.

Relatedly, participants reasoned about standardised vignettes rather than their own
symptoms. Judging an adapted case is not the same as facing a personal health concern: real
self-diagnosis carries emotional stakes that could either sharpen confirmation bias - people
may cling harder to a feared or hoped-for interpretation - or weaken AI reliance, if users
treat their own bodies as more authoritative than a generic recommendation. Our design
therefore has limited ecological validity for how confirmation bias and AI reliance operate
when the problem is one's own.

The study also used preconfigured AI responses rather than a live interactive system. This
improved experimental control and held content constant across participants, but reduced
ecological validity, as real symptom checkers respond dynamically to user input. The online
format added further technical and environmental variability - sessions ran on different
devices and in different settings, and although headphones were recommended, listening
conditions could not be controlled. A few participants restarted the study after technical
issues; their incomplete first sessions were discarded, as were all other incomplete sessions.
The unsupervised setting also affects response times: a small number of trials show very long
latencies (\rtLongN{} of \nTrials{} trials above 30\,s, maximum \rtMaxS\,s), most plausibly
because participants were briefly distracted (e.g., looked away from the screen) before
continuing. The app registered tab switches only during audio playback and cannot detect
off-screen attention, so such interruptions could not be identified reliably and no trials
were excluded; the response-time model was chosen to be robust to this right tail.

Our analyses characterise the confirmation effect at the level of the sample rather than
the individual. The mixed models and paired tests establish that, within participants,
contradiction triggered substantially more revision and confidence loss than confirmation;
but we did not classify participants by their general disposition to revise decisions or to
trust the AI. A participant who rarely changes their mind, or who distrusted the AI
throughout, can contribute to the aggregate pattern for reasons other than confirmation
bias, and cleanly separating such dispositional differences from a within-person
confirmation effect would require more trials per participant and explicit measures of
baseline revision tendency and trust.

The sample also skewed young (about \demoUnderThirtyfive\% under 35) and inexperienced with
these technologies (median prior experience \expScMed{} on the 1--7 scale), both of which
threaten generalisability in ways that could cut either direction: older adults may differ
in their familiarity with AI tools, baseline trust, and self-diagnosis habits, so the
effect's size - though probably not its direction - may not generalise to older populations;
and while inexperienced users have no calibrated basis for discounting a recommendation,
more experienced ones might weight AI advice differently, whether by trusting or scrutinising
it more. The magnitude of the effect may therefore vary across age and experience in ways our
sample cannot reveal.

Finally, and most relevant to our main question, the anthropomorphism gap we induced may
simply have been too small to move the bias: both voices were AI-generated and, though
distinguishable from each other, were rated as fairly artificial in absolute terms. A
genuinely human voice might produce a different result - a possibility our design cannot
rule out and that bears directly on our central research question.

These limitations point to clear next steps: stronger, differentiated voice manipulations -
ideally a genuinely human voice; medically validated cases with participants' own health
concerns, recovering the ecological validity vignettes sacrifice and letting AI direction be
crossed with correctness; and a real, dynamic AI chatbot in place of the pre-scripted
recommendations used here. Beyond voice, content-level interventions - explanations,
calibrated uncertainty, reflective prompts - may reduce confirmation bias and over-reliance,
since direction seems to matter more than delivery.

\section{Conclusion}

We investigated whether the voice design of an AI symptom checker shapes confirmation bias
in lay self-diagnosis. In a within-subjects online experiment ($N=\sampleN$), AI
recommendations exerted a strong, direction-dependent influence: participants changed their
diagnosis far more often and deliberated longer when the AI contradicted rather than
confirmed them, while their confidence rose substantially only after confirmation (all
$p<.001$). Contrary to our primary
hypothesis, this confirmation effect was statistically indistinguishable between an
anthropomorphic and a machine-like voice, even though the two voices were reliably
perceived as different. The presence and direction of an AI recommendation thus appear to
matter more for confirmation bias than the specific voice style used to deliver it, with the
design implications discussed above.

\section*{Open Science Statement}

All data, analysis code, study materials, and the study application are openly available at
\url{https://doi.org/10.5281/zenodo.22770605}. The archive contains the study application, the
trial-level and participant-level datasets in pseudonymised form together with a full codebook, the raw
export prior to cleaning (so that the exclusion of incomplete sessions can be verified),
the \textsf{R} scripts reproducing every value reported here, the twelve clinical vignettes
with their response options and spoken explanations, and the voice-perception instrument.
Data and materials are released under CC BY 4.0, code under the MIT licence.

Condition assignment was generated deterministically from a per-session seed, so the complete
assignment for a given session ID (block order, voice per block, case order, and AI
condition per trial) can be regenerated exactly from the published code. Session IDs
themselves are not published. Running the
deposited scripts reproduces the reported model estimates without modification.

Audio stimuli are not redistributed, because the terms of the two commercial
text-to-speech services used do not clearly permit it; the generation code, prompt
templates, and voice parameters are included so that the stimuli can be regenerated, and
the authors will provide the original files on request.

This study was not preregistered.

\begin{acks}
We thank all participants for taking part in the study. Claude (Anthropic) was used
to assist with developing the study software, and Gemini (Google) for literature
research, proofreading, and translation. All study content, design decisions,
analyses, and conclusions were created and verified by the authors.
\end{acks}

\bibliographystyle{ACM-Reference-Format-order}
\bibliography{references}


\begin{thebibliography}{34}


\ifx \showCODEN    \undefined \def \showCODEN     #1{\unskip}     \fi
\ifx \showISBNx    \undefined \def \showISBNx     #1{\unskip}     \fi
\ifx \showISBNxiii \undefined \def \showISBNxiii  #1{\unskip}     \fi
\ifx \showISSN     \undefined \def \showISSN      #1{\unskip}     \fi
\ifx \showLCCN     \undefined \def \showLCCN      #1{\unskip}     \fi
\ifx \shownote     \undefined \def \shownote      #1{#1}          \fi
\ifx \showarticletitle \undefined \def \showarticletitle #1{#1}   \fi
\ifx \showURL      \undefined \def \showURL       {\relax}        \fi
\providecommand\bibfield[2]{#2}
\providecommand\bibinfo[2]{#2}
\providecommand\natexlab[1]{#1}
\providecommand\showeprint[2][]{arXiv:#2}

\bibitem[Wallace et~al\mbox{.}(2022)]%
        {wallace2022}
\bibfield{author}{\bibinfo{person}{William Wallace}, \bibinfo{person}{Calvin
  Chan}, \bibinfo{person}{Swathikan Chidambaram}, \bibinfo{person}{Lydia
  Hanna}, \bibinfo{person}{Fahad~Mujtaba Iqbal}, \bibinfo{person}{Amish
  Acharya}, \bibinfo{person}{Pasha Normahani}, \bibinfo{person}{Hutan
  Ashrafian}, \bibinfo{person}{Sheraz~R. Markar}, \bibinfo{person}{Viknesh
  Sounderajah}, {and} \bibinfo{person}{Ara Darzi}.}
  \bibinfo{year}{2022}\natexlab{}.
\newblock \showarticletitle{The Diagnostic and Triage Accuracy of Digital and
  Online Symptom Checker Tools: A Systematic Review}.
\newblock \bibinfo{journal}{\emph{npj Digital Medicine}}  \bibinfo{volume}{5}
  (\bibinfo{year}{2022}), \bibinfo{pages}{118}.
\newblock
\showISSN{2398-6352}
\href{https://doi.org/10.1038/s41746-022-00667-w}{doi:\nolinkurl{10.1038/s41746-022-00667-w}}


\bibitem[Mentzou et~al\mbox{.}(2025)]%
        {mentzou2025}
\bibfield{author}{\bibinfo{person}{Aikaterini Mentzou}, \bibinfo{person}{Amy
  Rogers}, \bibinfo{person}{Edzia Carvalho}, \bibinfo{person}{Angela Daly},
  \bibinfo{person}{Maeve Malone}, {and} \bibinfo{person}{Xaroula Kerasidou}.}
  \bibinfo{year}{2025}\natexlab{}.
\newblock \showarticletitle{Artificial Intelligence in Digital Self-Diagnosis
  Tools: A Narrative Overview of Reviews}.
\newblock \bibinfo{journal}{\emph{Mayo Clinic Proceedings: Digital Health}}
  \bibinfo{volume}{3}, \bibinfo{number}{3} (\bibinfo{year}{2025}),
  \bibinfo{pages}{100242}.
\newblock
\showISSN{2949-7612}
\href{https://doi.org/10.1016/j.mcpdig.2025.100242}{doi:\nolinkurl{10.1016/j.mcpdig.2025.100242}}


\bibitem[K{\"u}cking et~al\mbox{.}(2024)]%
        {kuecking2024}
\bibfield{author}{\bibinfo{person}{Florian K{\"u}cking},
  \bibinfo{person}{Ursula H{\"u}bner}, \bibinfo{person}{Mareike Przysucha},
  \bibinfo{person}{Niels Hannemann}, \bibinfo{person}{Jan-Oliver Kutza},
  \bibinfo{person}{Maurice Moelleken}, \bibinfo{person}{Cornelia Erfurt-Berge},
  \bibinfo{person}{Joachim Dissemond}, \bibinfo{person}{Birgit Babitsch}, {and}
  \bibinfo{person}{Dorothee Busch}.} \bibinfo{year}{2024}\natexlab{}.
\newblock \showarticletitle{Automation Bias in {AI}-Decision Support: Results
  from an Empirical Study}.
\newblock \bibinfo{journal}{\emph{Studies in Health Technology and
  Informatics}}  \bibinfo{volume}{317} (\bibinfo{year}{2024}),
  \bibinfo{pages}{298--304}.
\newblock
\showISSN{0926-9630}
\href{https://doi.org/10.3233/SHTI240871}{doi:\nolinkurl{10.3233/SHTI240871}}


\bibitem[Qazi et~al\mbox{.}(2026)]%
        {qazi2026}
\bibfield{author}{\bibinfo{person}{Ihsan~Ayyub Qazi}, \bibinfo{person}{Ayesha
  Ali}, \bibinfo{person}{Asad~Ullah Khawaja}, \bibinfo{person}{Muhammad~Junaid
  Akhtar}, \bibinfo{person}{Ali~Zafar Sheikh}, {and}
  \bibinfo{person}{Muhammad~Hamad Alizai}.} \bibinfo{year}{2026}\natexlab{}.
\newblock \showarticletitle{Automation Bias in Large Language Model--Assisted
  Diagnostic Reasoning among Physicians Trained in {AI} Literacy: A Randomized
  Clinical Trial}.
\newblock \bibinfo{journal}{\emph{NEJM AI}} \bibinfo{volume}{3},
  \bibinfo{number}{5} (\bibinfo{year}{2026}), \bibinfo{pages}{AIoa2501001}.
\newblock
\href{https://doi.org/10.1056/AIoa2501001}{doi:\nolinkurl{10.1056/AIoa2501001}}


\bibitem[Rosbach et~al\mbox{.}(2025)]%
        {rosbach2025automation}
\bibfield{author}{\bibinfo{person}{Emely Rosbach}, \bibinfo{person}{Jonathan
  Ganz}, \bibinfo{person}{Jonas Ammeling}, \bibinfo{person}{Andreas Riener},
  {and} \bibinfo{person}{Marc Aubreville}.} \bibinfo{year}{2025}\natexlab{}.
\newblock \showarticletitle{Automation Bias in {AI}-Assisted Medical
  Decision-Making under Time Pressure in Computational Pathology}. In
  \bibinfo{booktitle}{\emph{Bildverarbeitung f{\"u}r die Medizin 2025}}
  \emph{(\bibinfo{series}{Informatik aktuell})}. \bibinfo{publisher}{Springer
  Vieweg}, \bibinfo{address}{Wiesbaden}, \bibinfo{pages}{129--134}.
\newblock
\href{https://doi.org/10.1007/978-3-658-47422-5_27}{doi:\nolinkurl{10.1007/978-3-658-47422-5_27}}


\bibitem[Nickerson(1998)]%
        {nickerson1998}
\bibfield{author}{\bibinfo{person}{Raymond~S. Nickerson}.}
  \bibinfo{year}{1998}\natexlab{}.
\newblock \showarticletitle{Confirmation Bias: A Ubiquitous Phenomenon in Many
  Guises}.
\newblock \bibinfo{journal}{\emph{Review of General Psychology}}
  \bibinfo{volume}{2}, \bibinfo{number}{2} (\bibinfo{year}{1998}),
  \bibinfo{pages}{175--220}.
\newblock
\showISSN{1089-2680}
\href{https://doi.org/10.1037/1089-2680.2.2.175}{doi:\nolinkurl{10.1037/1089-2680.2.2.175}}


\bibitem[Rosbach et~al\mbox{.}(2025)]%
        {rosbach2024}
\bibfield{author}{\bibinfo{person}{Emely Rosbach}, \bibinfo{person}{Jonas
  Ammeling}, \bibinfo{person}{Sebastian Kr{\"u}gel}, \bibinfo{person}{Angelika
  Kie{\ss}ig}, \bibinfo{person}{Alexis Fritz}, \bibinfo{person}{Jonathan Ganz},
  \bibinfo{person}{Chlo{\'e} Puget}, \bibinfo{person}{Taryn Donovan},
  \bibinfo{person}{Andrea Klang}, \bibinfo{person}{Maximilian~C. K{\"o}ller},
  \bibinfo{person}{Pompei Bolfa}, \bibinfo{person}{Marco Tecilla},
  \bibinfo{person}{Daniela Denk}, \bibinfo{person}{Matti Kiupel},
  \bibinfo{person}{Georgios Paraschou}, \bibinfo{person}{Mun~Keong Kok},
  \bibinfo{person}{Alexander F.~H. Haake}, \bibinfo{person}{Ronald~R. de
  Krijger}, \bibinfo{person}{Andreas F.-P. Sonnen}, \bibinfo{person}{Tanit
  Kasantikul}, \bibinfo{person}{Gerry~M. Dorrestein},
  \bibinfo{person}{Rebecca~C. Smedley}, \bibinfo{person}{Nikolas Stathonikos},
  \bibinfo{person}{Matthias Uhl}, \bibinfo{person}{Christof~A. Bertram},
  \bibinfo{person}{Andreas Riener}, {and} \bibinfo{person}{Marc Aubreville}.}
  \bibinfo{year}{2025}\natexlab{}.
\newblock \showarticletitle{``When Two Wrongs Don't Make a Right'' -- Examining
  Confirmation Bias and the Role of Time Pressure During Human-{AI}
  Collaboration in Computational Pathology}. In
  \bibinfo{booktitle}{\emph{Proceedings of the 2025 CHI Conference on Human
  Factors in Computing Systems}} \emph{(\bibinfo{series}{CHI '25})}.
  \bibinfo{publisher}{ACM}, \bibinfo{address}{New York, NY, USA},
  \bibinfo{numpages}{18}~pages.
\newblock
\href{https://doi.org/10.1145/3706598.3713319}{doi:\nolinkurl{10.1145/3706598.3713319}}


\bibitem[Seaborn et~al\mbox{.}(2021)]%
        {seaborn2021}
\bibfield{author}{\bibinfo{person}{Katie Seaborn}, \bibinfo{person}{Norihisa~P.
  Miyake}, \bibinfo{person}{Peter Pennefather}, {and} \bibinfo{person}{Mihoko
  Otake-Matsuura}.} \bibinfo{year}{2021}\natexlab{}.
\newblock \showarticletitle{Voice in Human--Agent Interaction: A Survey}.
\newblock \bibinfo{journal}{\emph{Comput. Surveys}} \bibinfo{volume}{54},
  \bibinfo{number}{4} (\bibinfo{year}{2021}), \bibinfo{pages}{1--43}.
\newblock
\href{https://doi.org/10.1145/3386867}{doi:\nolinkurl{10.1145/3386867}}


\bibitem[Zhan et~al\mbox{.}(2024)]%
        {zhan2024}
\bibfield{author}{\bibinfo{person}{Xiao Zhan}, \bibinfo{person}{Noura Abdi},
  \bibinfo{person}{William Seymour}, {and} \bibinfo{person}{Jose Such}.}
  \bibinfo{year}{2024}\natexlab{}.
\newblock \showarticletitle{Healthcare Voice {AI} Assistants: Factors
  Influencing Trust and Intention to Use}.
\newblock \bibinfo{journal}{\emph{Proceedings of the ACM on Human-Computer
  Interaction}} \bibinfo{volume}{8}, \bibinfo{number}{CSCW1}, Article
  \bibinfo{articleno}{62} (\bibinfo{year}{2024}), \bibinfo{numpages}{37}~pages.
\newblock
\showISSN{2573-0142}
\href{https://doi.org/10.1145/3637339}{doi:\nolinkurl{10.1145/3637339}}


\bibitem[Milford et~al\mbox{.}(2025)]%
        {milford2025}
\bibfield{author}{\bibinfo{person}{Stephen~R. Milford}, \bibinfo{person}{Emma
  Herger}, \bibinfo{person}{Johanna Eichinger}, {and}
  \bibinfo{person}{Bernice~S. Elger}.} \bibinfo{year}{2025}\natexlab{}.
\newblock \showarticletitle{Promoting Xenomorphic Patient-Facing {AIs}: The
  Case Against Anthropomorphism in Medical {AIs}}.
\newblock \bibinfo{journal}{\emph{npj Digital Medicine}}  \bibinfo{volume}{8}
  (\bibinfo{year}{2025}), \bibinfo{pages}{667}.
\newblock
\showISSN{2398-6352}
\href{https://doi.org/10.1038/s41746-025-02046-7}{doi:\nolinkurl{10.1038/s41746-025-02046-7}}


\bibitem[Bashkirova and Krpan(2024)]%
        {bashkirova2024}
\bibfield{author}{\bibinfo{person}{Anna Bashkirova} {and}
  \bibinfo{person}{Dario Krpan}.} \bibinfo{year}{2024}\natexlab{}.
\newblock \showarticletitle{Confirmation Bias in {AI}-Assisted Decision-Making:
  {AI} Triage Recommendations Congruent with Expert Judgments Increase
  Psychologist Trust and Recommendation Acceptance}.
\newblock \bibinfo{journal}{\emph{Computers in Human Behavior: Artificial
  Humans}} \bibinfo{volume}{2}, \bibinfo{number}{1} (\bibinfo{year}{2024}),
  \bibinfo{pages}{100066}.
\newblock
\href{https://doi.org/10.1016/j.chbah.2024.100066}{doi:\nolinkurl{10.1016/j.chbah.2024.100066}}


\bibitem[Schultze et~al\mbox{.}(2015)]%
        {schultze2015}
\bibfield{author}{\bibinfo{person}{Thomas Schultze},
  \bibinfo{person}{Anne-Fernandine Rakotoarisoa}, {and} \bibinfo{person}{Stefan
  Schulz-Hardt}.} \bibinfo{year}{2015}\natexlab{}.
\newblock \showarticletitle{Effects of Distance Between Initial Estimates and
  Advice on Advice Utilization}.
\newblock \bibinfo{journal}{\emph{Judgment and Decision Making}}
  \bibinfo{volume}{10}, \bibinfo{number}{2} (\bibinfo{year}{2015}),
  \bibinfo{pages}{144--171}.
\newblock
\href{https://doi.org/10.1017/S1930297500003922}{doi:\nolinkurl{10.1017/S1930297500003922}}


\bibitem[Yaniv et~al\mbox{.}(2009)]%
        {yaniv2009}
\bibfield{author}{\bibinfo{person}{Ilan Yaniv}, \bibinfo{person}{Shoham
  Choshen-Hillel}, {and} \bibinfo{person}{Maxim Milyavsky}.}
  \bibinfo{year}{2009}\natexlab{}.
\newblock \showarticletitle{Spurious Consensus and Opinion Revision: Why Might
  People Be More Confident in Their Less Accurate Judgments?}
\newblock \bibinfo{journal}{\emph{Journal of Experimental Psychology: Learning,
  Memory, and Cognition}} \bibinfo{volume}{35}, \bibinfo{number}{2}
  (\bibinfo{year}{2009}), \bibinfo{pages}{558--563}.
\newblock
\href{https://doi.org/10.1037/a0014589}{doi:\nolinkurl{10.1037/a0014589}}


\bibitem[Kappes et~al\mbox{.}(2020)]%
        {kappes2020}
\bibfield{author}{\bibinfo{person}{Andreas Kappes}, \bibinfo{person}{Ann~H.
  Harvey}, \bibinfo{person}{Terry Lohrenz}, \bibinfo{person}{P.~Read Montague},
  {and} \bibinfo{person}{Tali Sharot}.} \bibinfo{year}{2020}\natexlab{}.
\newblock \showarticletitle{Confirmation Bias in the Utilization of Others'
  Opinion Strength}.
\newblock \bibinfo{journal}{\emph{Nature Neuroscience}} \bibinfo{volume}{23},
  \bibinfo{number}{1} (\bibinfo{year}{2020}), \bibinfo{pages}{130--137}.
\newblock
\href{https://doi.org/10.1038/s41593-019-0549-2}{doi:\nolinkurl{10.1038/s41593-019-0549-2}}


\bibitem[Vodrahalli et~al\mbox{.}(2022)]%
        {vodrahalli2022}
\bibfield{author}{\bibinfo{person}{Kailas Vodrahalli}, \bibinfo{person}{Roxana
  Daneshjou}, \bibinfo{person}{Tobias Gerstenberg}, {and}
  \bibinfo{person}{James Zou}.} \bibinfo{year}{2022}\natexlab{}.
\newblock \showarticletitle{Do Humans Trust Advice More if it Comes from {AI}?
  An Analysis of Human-{AI} Interactions}. In
  \bibinfo{booktitle}{\emph{Proceedings of the 2022 AAAI/ACM Conference on AI,
  Ethics, and Society}} \emph{(\bibinfo{series}{AIES '22})}.
  \bibinfo{publisher}{ACM}, \bibinfo{address}{New York, NY, USA},
  \bibinfo{pages}{763--777}.
\newblock
\href{https://doi.org/10.1145/3514094.3534150}{doi:\nolinkurl{10.1145/3514094.3534150}}


\bibitem[Lu et~al\mbox{.}(2024)]%
        {lu2024}
\bibfield{author}{\bibinfo{person}{Zhuoran Lu}, \bibinfo{person}{Dakuo Wang},
  {and} \bibinfo{person}{Ming Yin}.} \bibinfo{year}{2024}\natexlab{}.
\newblock \showarticletitle{Does More Advice Help? The Effects of Second
  Opinions in {AI}-Assisted Decision Making}.
\newblock \bibinfo{journal}{\emph{Proceedings of the ACM on Human-Computer
  Interaction}} \bibinfo{volume}{8}, \bibinfo{number}{CSCW1}
  (\bibinfo{year}{2024}), \bibinfo{pages}{1--31}.
\newblock
\href{https://doi.org/10.1145/3653708}{doi:\nolinkurl{10.1145/3653708}}


\bibitem[Rastogi et~al\mbox{.}(2022)]%
        {rastogi2022}
\bibfield{author}{\bibinfo{person}{Charvi Rastogi}, \bibinfo{person}{Yunfeng
  Zhang}, \bibinfo{person}{Dennis Wei}, \bibinfo{person}{Kush~R. Varshney},
  \bibinfo{person}{Amit Dhurandhar}, {and} \bibinfo{person}{Richard Tomsett}.}
  \bibinfo{year}{2022}\natexlab{}.
\newblock \showarticletitle{Deciding Fast and Slow: The Role of Cognitive
  Biases in {AI}-Assisted Decision-Making}.
\newblock \bibinfo{journal}{\emph{Proceedings of the ACM on Human-Computer
  Interaction}} \bibinfo{volume}{6}, \bibinfo{number}{CSCW1}
  (\bibinfo{year}{2022}), \bibinfo{pages}{1--22}.
\newblock
\href{https://doi.org/10.1145/3512930}{doi:\nolinkurl{10.1145/3512930}}


\bibitem[Yu and Chen(2024)]%
        {yu2024}
\bibfield{author}{\bibinfo{person}{Shulan Yu} {and} \bibinfo{person}{Tianyue
  Chen}.} \bibinfo{year}{2024}\natexlab{}.
\newblock \showarticletitle{Understanding Older Adults' Acceptance of Chatbots
  in Healthcare Delivery: An Extended {UTAUT} Model}.
\newblock \bibinfo{journal}{\emph{Frontiers in Public Health}}
  \bibinfo{volume}{12} (\bibinfo{year}{2024}), \bibinfo{pages}{1435329}.
\newblock
\showISSN{2296-2565}
\href{https://doi.org/10.3389/fpubh.2024.1435329}{doi:\nolinkurl{10.3389/fpubh.2024.1435329}}


\bibitem[Zada et~al\mbox{.}(2025)]%
        {zada2025}
\bibfield{author}{\bibinfo{person}{Troy Zada}, \bibinfo{person}{Natalie Tam},
  \bibinfo{person}{Francois Barnard}, \bibinfo{person}{Marlize van Sittert},
  \bibinfo{person}{Venkat Bhat}, {and} \bibinfo{person}{Sirisha Rambhatla}.}
  \bibinfo{year}{2025}\natexlab{}.
\newblock \showarticletitle{Medical Misinformation in {AI}-Assisted
  Self-Diagnosis: Development of a Method ({EvalPrompt}) for Analyzing Large
  Language Models}.
\newblock \bibinfo{journal}{\emph{JMIR Formative Research}}
  \bibinfo{volume}{9} (\bibinfo{year}{2025}), \bibinfo{pages}{e66207}.
\newblock
\showISSN{2561-326X}
\href{https://doi.org/10.2196/66207}{doi:\nolinkurl{10.2196/66207}}


\bibitem[You and Gui(2021)]%
        {you2021}
\bibfield{author}{\bibinfo{person}{Yue You} {and} \bibinfo{person}{Xinning
  Gui}.} \bibinfo{year}{2021}\natexlab{}.
\newblock \bibinfo{title}{Self-Diagnosis through {AI}-Enabled Chatbot-Based
  Symptom Checkers: User Experiences and Design Considerations}.
\newblock \bibinfo{howpublished}{arXiv:2101.04796}.
\newblock
\href{https://doi.org/10.48550/arXiv.2101.04796}{doi:\nolinkurl{10.48550/arXiv.2101.04796}}


\bibitem[Tu et~al\mbox{.}(2025)]%
        {tu2025}
\bibfield{author}{\bibinfo{person}{Tao Tu}, \bibinfo{person}{Anil Palepu},
  \bibinfo{person}{Mike Schaekermann}, \bibinfo{person}{Khaled Saab},
  \bibinfo{person}{Jan Freyberg}, \bibinfo{person}{Ryutaro Tanno},
  \bibinfo{person}{Amy Wang}, \bibinfo{person}{Brenna Li},
  \bibinfo{person}{Mohamed Amin}, \bibinfo{person}{Yong Cheng},
  \bibinfo{person}{Elahe Vedadi}, \bibinfo{person}{Nenad Tomasev},
  \bibinfo{person}{Shekoofeh Azizi}, \bibinfo{person}{Karan Singhal},
  \bibinfo{person}{Le Hou}, \bibinfo{person}{Albert Webson},
  \bibinfo{person}{Kavita Kulkarni}, \bibinfo{person}{S.~Sara Mahdavi},
  \bibinfo{person}{Christopher Semturs}, \bibinfo{person}{Juraj Gottweis},
  \bibinfo{person}{Joelle Barral}, \bibinfo{person}{Katherine Chou},
  \bibinfo{person}{Greg~S. Corrado}, \bibinfo{person}{Yossi Matias},
  \bibinfo{person}{Alan Karthikesalingam}, {and} \bibinfo{person}{Vivek
  Natarajan}.} \bibinfo{year}{2025}\natexlab{}.
\newblock \showarticletitle{Towards Conversational Diagnostic Artificial
  Intelligence}.
\newblock \bibinfo{journal}{\emph{Nature}}  \bibinfo{volume}{642}
  (\bibinfo{year}{2025}), \bibinfo{pages}{442--450}.
\newblock
\showISSN{1476-4687}
\href{https://doi.org/10.1038/s41586-025-08866-7}{doi:\nolinkurl{10.1038/s41586-025-08866-7}}


\bibitem[Cross et~al\mbox{.}(2024)]%
        {cross2024}
\bibfield{author}{\bibinfo{person}{James~L. Cross}, \bibinfo{person}{Michael~A.
  Choma}, {and} \bibinfo{person}{John~A. Onofrey}.}
  \bibinfo{year}{2024}\natexlab{}.
\newblock \showarticletitle{Bias in Medical {AI}: Implications for Clinical
  Decision-Making}.
\newblock \bibinfo{journal}{\emph{PLOS Digital Health}} \bibinfo{volume}{3},
  \bibinfo{number}{11} (\bibinfo{year}{2024}), \bibinfo{pages}{e0000651}.
\newblock
\showISSN{2767-3170}
\href{https://doi.org/10.1371/journal.pdig.0000651}{doi:\nolinkurl{10.1371/journal.pdig.0000651}}


\bibitem[Kirkland et~al\mbox{.}(2022)]%
        {kirkland2022}
\bibfield{author}{\bibinfo{person}{Ambika Kirkland}, \bibinfo{person}{Harm
  Lameris}, \bibinfo{person}{{\'E}va Sz{\'e}kely}, {and}
  \bibinfo{person}{Joakim Gustafson}.} \bibinfo{year}{2022}\natexlab{}.
\newblock \showarticletitle{Where's the Uh, Hesitation? The Interplay between
  Filled Pause Location, Speech Rate and Fundamental Frequency in Perception of
  Confidence}. In \bibinfo{booktitle}{\emph{Proceedings of Interspeech 2022}}.
  \bibinfo{publisher}{International Speech Communication Association},
  \bibinfo{address}{Baixas}, \bibinfo{pages}{4990--4994}.
\newblock
\showISSN{2958-1796}
\href{https://doi.org/10.21437/Interspeech.2022-10973}{doi:\nolinkurl{10.21437/Interspeech.2022-10973}}


\bibitem[Dubiel et~al\mbox{.}(2024)]%
        {dubiel2024}
\bibfield{author}{\bibinfo{person}{Mateusz Dubiel}, \bibinfo{person}{Anastasia
  Sergeeva}, {and} \bibinfo{person}{Luis~A. Leiva}.}
  \bibinfo{year}{2024}\natexlab{}.
\newblock \showarticletitle{Impact of Voice Fidelity on Decision Making: A
  Potential Dark Pattern?}. In \bibinfo{booktitle}{\emph{Proceedings of the
  29th International Conference on Intelligent User Interfaces (IUI '24)}}.
  \bibinfo{publisher}{Association for Computing Machinery},
  \bibinfo{address}{New York, NY, USA}, \bibinfo{pages}{181--194}.
\newblock
\showISBNx{979-8-4007-0508-3}
\href{https://doi.org/10.1145/3640543.3645202}{doi:\nolinkurl{10.1145/3640543.3645202}}


\bibitem[Abercrombie et~al\mbox{.}(2021)]%
        {abercrombie2021}
\bibfield{author}{\bibinfo{person}{Gavin Abercrombie}, \bibinfo{person}{Amanda
  Cercas~Curry}, \bibinfo{person}{Mugdha Pandya}, {and} \bibinfo{person}{Verena
  Rieser}.} \bibinfo{year}{2021}\natexlab{}.
\newblock \showarticletitle{Alexa, Google, Siri: What Are Your Pronouns? Gender
  and Anthropomorphism in the Design and Perception of Conversational
  Assistants}. In \bibinfo{booktitle}{\emph{Proceedings of the 3rd Workshop on
  Gender Bias in Natural Language Processing}}. \bibinfo{publisher}{Association
  for Computational Linguistics}, \bibinfo{address}{Online},
  \bibinfo{pages}{24--33}.
\newblock
\href{https://doi.org/10.18653/v1/2021.gebnlp-1.4}{doi:\nolinkurl{10.18653/v1/2021.gebnlp-1.4}}


\bibitem[{Dhivyesh R K}(2023)]%
        {dhivyeshrk2023}
\bibfield{author}{\bibinfo{person}{{Dhivyesh R K}}.}
  \bibinfo{year}{2023}\natexlab{}.
\newblock \bibinfo{title}{Disease-Symptom Dataset}.
\newblock \bibinfo{howpublished}{Kaggle}.
\newblock
\urldef\tempurl%
\url{https://www.kaggle.com/datasets/dhivyeshrk/diseases-and-symptoms-dataset}
\showURL{%
\tempurl}
\newblock
\shownote{Dataset file:
  Final\_Augmented\_dataset\_Diseases\_and\_Symptoms.csv}.


\bibitem[Bartneck et~al\mbox{.}(2009)]%
        {bartneck2009}
\bibfield{author}{\bibinfo{person}{Christoph Bartneck}, \bibinfo{person}{Dana
  Kuli{\'c}}, \bibinfo{person}{Elizabeth Croft}, {and} \bibinfo{person}{Susana
  Zoghbi}.} \bibinfo{year}{2009}\natexlab{}.
\newblock \showarticletitle{Measurement Instruments for the Anthropomorphism,
  Animacy, Likeability, Perceived Intelligence, and Perceived Safety of
  Robots}.
\newblock \bibinfo{journal}{\emph{International Journal of Social Robotics}}
  \bibinfo{volume}{1}, \bibinfo{number}{1} (\bibinfo{year}{2009}),
  \bibinfo{pages}{71--81}.
\newblock
\href{https://doi.org/10.1007/s12369-008-0001-3}{doi:\nolinkurl{10.1007/s12369-008-0001-3}}


\bibitem[K{\"u}hne et~al\mbox{.}(2020)]%
        {kuehne2020}
\bibfield{author}{\bibinfo{person}{Katharina K{\"u}hne},
  \bibinfo{person}{Martin~H. Fischer}, {and} \bibinfo{person}{Yuefang Zhou}.}
  \bibinfo{year}{2020}\natexlab{}.
\newblock \showarticletitle{The Human Takes It All: Humanlike Synthesized
  Voices Are Perceived as Less Eerie and More Likable. Evidence From a
  Subjective Ratings Study}.
\newblock \bibinfo{journal}{\emph{Frontiers in Neurorobotics}}
  \bibinfo{volume}{14} (\bibinfo{year}{2020}), \bibinfo{pages}{593732}.
\newblock
\href{https://doi.org/10.3389/fnbot.2020.593732}{doi:\nolinkurl{10.3389/fnbot.2020.593732}}


\bibitem[Fiske et~al\mbox{.}(2007)]%
        {fiske2007}
\bibfield{author}{\bibinfo{person}{Susan~T. Fiske}, \bibinfo{person}{Amy J.~C.
  Cuddy}, {and} \bibinfo{person}{Peter Glick}.}
  \bibinfo{year}{2007}\natexlab{}.
\newblock \showarticletitle{Universal Dimensions of Social Cognition: Warmth
  and Competence}.
\newblock \bibinfo{journal}{\emph{Trends in Cognitive Sciences}}
  \bibinfo{volume}{11}, \bibinfo{number}{2} (\bibinfo{year}{2007}),
  \bibinfo{pages}{77--83}.
\newblock
\href{https://doi.org/10.1016/j.tics.2006.11.005}{doi:\nolinkurl{10.1016/j.tics.2006.11.005}}


\bibitem[Bates et~al\mbox{.}(2015)]%
        {bates2015}
\bibfield{author}{\bibinfo{person}{Douglas Bates}, \bibinfo{person}{Martin
  M{\"a}chler}, \bibinfo{person}{Ben Bolker}, {and} \bibinfo{person}{Steve
  Walker}.} \bibinfo{year}{2015}\natexlab{}.
\newblock \showarticletitle{Fitting Linear Mixed-Effects Models Using {lme4}}.
\newblock \bibinfo{journal}{\emph{Journal of Statistical Software}}
  \bibinfo{volume}{67}, \bibinfo{number}{1} (\bibinfo{year}{2015}),
  \bibinfo{pages}{1--48}.
\newblock
\href{https://doi.org/10.18637/jss.v067.i01}{doi:\nolinkurl{10.18637/jss.v067.i01}}


\bibitem[Koller(2016)]%
        {koller2016}
\bibfield{author}{\bibinfo{person}{Manuel Koller}.}
  \bibinfo{year}{2016}\natexlab{}.
\newblock \showarticletitle{{robustlmm}: An {R} Package for Robust Estimation
  of Linear Mixed-Effects Models}.
\newblock \bibinfo{journal}{\emph{Journal of Statistical Software}}
  \bibinfo{volume}{75}, \bibinfo{number}{6} (\bibinfo{year}{2016}),
  \bibinfo{pages}{1--24}.
\newblock
\href{https://doi.org/10.18637/jss.v075.i06}{doi:\nolinkurl{10.18637/jss.v075.i06}}


\bibitem[Lo and Andrews(2015)]%
        {lo2015}
\bibfield{author}{\bibinfo{person}{Steson Lo} {and} \bibinfo{person}{Sally
  Andrews}.} \bibinfo{year}{2015}\natexlab{}.
\newblock \showarticletitle{To Transform or Not to Transform: Using Generalized
  Linear Mixed Models to Analyse Reaction Time Data}.
\newblock \bibinfo{journal}{\emph{Frontiers in Psychology}}
  \bibinfo{volume}{6} (\bibinfo{year}{2015}), \bibinfo{pages}{1171}.
\newblock
\href{https://doi.org/10.3389/fpsyg.2015.01171}{doi:\nolinkurl{10.3389/fpsyg.2015.01171}}


\bibitem[Brooks et~al\mbox{.}(2017)]%
        {brooks2017}
\bibfield{author}{\bibinfo{person}{Mollie~E. Brooks}, \bibinfo{person}{Kasper
  Kristensen}, \bibinfo{person}{Koen~J. van Benthem}, \bibinfo{person}{Arni
  Magnusson}, \bibinfo{person}{Casper~W. Berg}, \bibinfo{person}{Anders
  Nielsen}, \bibinfo{person}{Hans~J. Skaug}, \bibinfo{person}{Martin
  M{\"a}chler}, {and} \bibinfo{person}{Benjamin~M. Bolker}.}
  \bibinfo{year}{2017}\natexlab{}.
\newblock \showarticletitle{{glmmTMB} Balances Speed and Flexibility Among
  Packages for Zero-inflated Generalized Linear Mixed Modeling}.
\newblock \bibinfo{journal}{\emph{The R Journal}} \bibinfo{volume}{9},
  \bibinfo{number}{2} (\bibinfo{year}{2017}), \bibinfo{pages}{378--400}.
\newblock
\href{https://doi.org/10.32614/RJ-2017-066}{doi:\nolinkurl{10.32614/RJ-2017-066}}


\bibitem[Rollwage et~al\mbox{.}(2020)]%
        {rollwage2020}
\bibfield{author}{\bibinfo{person}{Max Rollwage}, \bibinfo{person}{Alisa
  Loosen}, \bibinfo{person}{Tobias~U. Hauser}, \bibinfo{person}{Rani Moran},
  \bibinfo{person}{Raymond~J. Dolan}, {and} \bibinfo{person}{Stephen~M.
  Fleming}.} \bibinfo{year}{2020}\natexlab{}.
\newblock \showarticletitle{Confidence Drives a Neural Confirmation Bias}.
\newblock \bibinfo{journal}{\emph{Nature Communications}} \bibinfo{volume}{11},
  \bibinfo{number}{1} (\bibinfo{year}{2020}), \bibinfo{pages}{2634}.
\newblock
\href{https://doi.org/10.1038/s41467-020-16278-6}{doi:\nolinkurl{10.1038/s41467-020-16278-6}}


\end{thebibliography}

\appendix
\section{Residual Diagnostics}
\label{app:residuals}

\begin{figure}[htbp]
  \centering
  \begin{minipage}[t]{0.48\columnwidth}
    \centering
    \includegraphics[width=\linewidth]{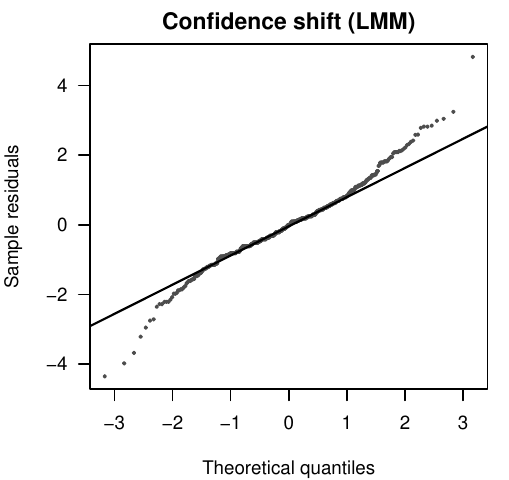}
    \par\vspace{2pt}
    \footnotesize (a) Confidence shift
  \end{minipage}\hfill
  \begin{minipage}[t]{0.48\columnwidth}
    \centering
    \includegraphics[width=\linewidth]{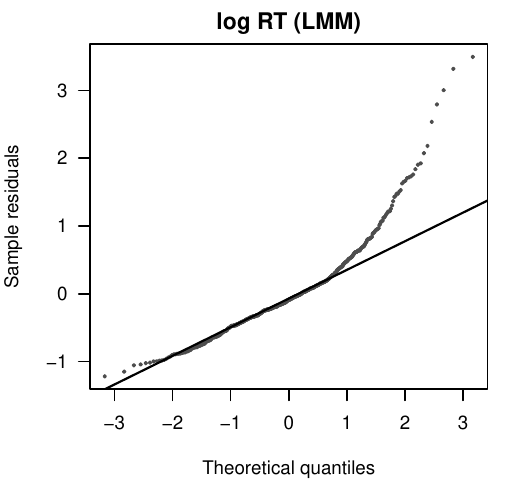}
    \par\vspace{2pt}
    \footnotesize (b) log response time
  \end{minipage}
  \caption{Normal Q--Q plots of the LMM conditional residuals for the
    confidence-shift outcome (a) and the log response-time outcome (b).
    Points would fall on the diagonal under normality; both show
    clear departures in the tails, and the response-time residuals
    remain right-skewed.}
  \Description{Two quantile-quantile plots of model residuals. Both follow the diagonal through the centre with deviations in the tails, more pronounced for the response-time model.}
  \label{fig:qq}
\end{figure}

To choose appropriate models for the two continuous outcomes, we first
fitted Gaussian linear mixed models (LMMs) to the confidence shift and to
log-transformed response times and inspected their residuals.
Figure~\ref{fig:qq} shows normal quantile--quantile plots of the
conditional residuals of these two models.

Formal Shapiro--Wilk tests reject normality for both outcomes (confidence shift
$W=\shapiroShiftW$, $p\shapiroShiftP$; log response time $W=\shapiroRtW$, $p\shapiroRtP$),
and the Q--Q plots show heavy tails for the confidence shift and a remaining right skew for
response time despite the log transformation. Because the Shapiro--Wilk test is very
sensitive at this sample size and the confidence ratings are bounded, we use the test only
together with the Q--Q plots, as evidence that the normality assumption is not met. This
motivated the primary models reported in the main text: a robust LMM for the confidence shift
and a Gamma GLMM with log link on untransformed seconds for response time. The
diagnosis-change outcome is analysed with a binomial GLMM, which makes no residual-normality
assumption.

\section{Sensitivity Analysis: Gaussian LMMs}
\label{app:sensitivity}

Table~\ref{tab:sensitivity} compares the AI main effect and the AI\,$\times$\,voice
interaction of the primary models with those of Gaussian LMMs fitted to the confidence shift
and to log-transformed response times. The conclusions are the same under both model
families: the AI effect is large and significant, and the interaction is small and
non-significant.

\begin{table}
  \centering
  \caption{Sensitivity analysis: AI main effect and AI\,$\times$\,Voice interaction
    (sum-coded coefficients \(b\) with Wald 95\,\% CIs) in the primary models and in
    Gaussian LMMs, which assume normally distributed residuals.}
  \label{tab:sensitivity}
  \small
  \setlength{\tabcolsep}{3pt}
  \begin{tabular}{lllr}
    \toprule
    Model & Term & \(b\) [95\%~CI] & \(p\)\\
    \midrule
    Shift, robust LMM & AI & $0.505$ [$0.434$, $0.577$] & ${<}.001$\\
    Shift, Gaussian LMM & AI & $0.523$ [$0.441$, $0.605$] & ${<}.001$\\
    Shift, robust LMM & AI\,$\times$\,V & $0.011$ [$-0.061$, $0.083$] & $.759$\\
    Shift, Gaussian LMM & AI\,$\times$\,V & $-0.005$ [$-0.087$, $0.077$] & $.912$\\
    \midrule
    Time, Gamma GLMM & AI & $-0.299$ [$-0.351$, $-0.247$] & ${<}.001$\\
    Time, LMM on log & AI & $-0.250$ [$-0.299$, $-0.202$] & ${<}.001$\\
    Time, Gamma GLMM & AI\,$\times$\,V & $-0.021$ [$-0.071$, $0.030$] & $.420$\\
    Time, LMM on log & AI\,$\times$\,V & $-0.002$ [$-0.050$, $0.047$] & $.937$\\
    \bottomrule
  \end{tabular}
\end{table}

\section{Participant-Level Tests}
\label{app:participant}

To check that the trial-level results do not depend on the mixed-model assumptions, we also
analysed data aggregated per participant. For diagnosis change, each participant's
proportion of changed diagnoses was computed in each of the four AI\,$\times$\,voice cells
(two trials per cell) and compared with a Friedman test with participant as the blocking
factor; the cells differed significantly ($\chi^2(3)=\friedmanPickChi$, $p\friedmanPickP$,
Kendall's $W=\kendallPickW$). Additionally, a McNemar test compared, within participants,
whether a person changed their diagnosis on at least one trial under contradiction and under
confirmation: \changeRefutePart{} of participants did so at least once under contradiction
versus \changeConfirmPart{} under confirmation ($\chi^2=\mcnemarChi$, $p\mcnemarP$). For the
confidence shift, a Friedman test on the per-participant cell means likewise showed a
significant difference between the four cells ($\chi^2(3)=\friedmanChi$, $p\friedmanP$,
Kendall's $W=\kendallW$). Holm-corrected Wilcoxon signed-rank tests trace both Friedman results to the AI condition:
the confirm and contradict cells differed within each voice (all $p\posthocAiMaxP$), whereas
the two voices did not differ within either AI condition (all $p\posthocVoiceMinP$ or
larger).

\end{document}